\documentclass[apsrev4-1, prb, twocolumn, footinbib, superscriptaddress, floatfix, bibliography]{revtex4-2}

\usepackage[dvipdfmx]{graphicx}
\usepackage{amsmath,amsbsy,amssymb}
\usepackage{bm}
\usepackage{mathrsfs}
\usepackage{textgreek}
\usepackage{mathrsfs}
\usepackage{multirow}
\usepackage{physics}
\usepackage{comment}

\usepackage{color}

\usepackage[colorlinks,bookmarks=false,citecolor=blue,linkcolor=red,urlcolor=blue]{hyperref}

\usepackage[caption=false]{subfig}	
\newcommand{\diff}{\mathrm{d}}

\newcommand{\imu}{\mathrm{i}}
\newcommand{\epn}{\mathrm{e}}

\newcommand{\dg}{\dagger}
\newcommand{\la}{\langle}
\newcommand{\ra}{\rangle}
\newcommand{\al}{\alpha}
\newcommand{\sg}{\sigma}
\newcommand{\gm}{\gamma}
\newcommand{\ep}{\varepsilon}

\begin{document}

\title{
 Electron-phonon-coupled Langevin dynamics for strongly-correlated insulators
}

\author{Rico Pohle}
\affiliation{Faculty of Science, 
Shizuoka University, Shizuoka 422-8529, Japan}
\affiliation{Institute for Materials Research,
Tohoku University, Sendai, Miyagi 980-8577, Japan}

\author{Yukitoshi Motome}
\affiliation{Department of Applied Physics, 
University of Tokyo, Hongo, Tokyo 113-8656, Japan}

\author{Terumasa Tadano}
\affiliation{Research Center for Magnetic and Spintronic Materials, 
National Institute for Materials Science, Tsukuba 305-0047, Japan}

\author{Shintaro Hoshino}
\affiliation{Department of Physics, 
Chiba University, Chiba 263-8522, Japan}
\affiliation{Department of Physics, 
Saitama University, Saitama 338-8570, Japan}

\date{\today}

\begin{abstract}

The Landau-Lifshitz-Gilbert (LLG) equations are widely used to study 
spin dynamics in Mott insulators. 
However, because energy damping 
is typically introduced phenomenologically, their 
validity for describing nonequilibrium processes and 
their connection to the microscopic origin of dissipation 
in real materials remains unclear.
In this paper, we derive 
generalized stochastic LLG equations from first 
principles for spin-orbital coupled Mott insulators, explicitly 
incorporating the coupling 
between electronic degrees of freedom and lattice vibrations. 
Our approach is based on a 
path-integral formalism formulated along the 
Keldysh contour, which naturally accounts for dissipation and thermal fluctuations
through interactions with a phonon bath and emergent stochastic noise.
We benchmark our theoretical framework by numerically integrating the 
equations of motion for a two-orbital
spin chain coupled to Einstein phonons. 
The resulting energy relaxation mimics realistic cooling dynamics, exhibits 
nontrivial transient behavior during thermalization, and accurately 
reproduces thermodynamic properties upon equilibration.
We further demonstrate how electron–phonon coupling induces 
hybridization between electronic and phononic modes in the excitation spectrum
and show that the conventional LLG equations 
are recovered as a limiting case of our microscopic theory.
These results establish a robust and reliable framework 
for capturing dissipative spin dynamics in strongly correlated systems, 
both in and out of equilibrium.

\end{abstract}

\maketitle

\section{Introduction}
\label{sec:Introduction}

\begin{figure}[t]
    \centering
    \includegraphics[width=0.5\textwidth]{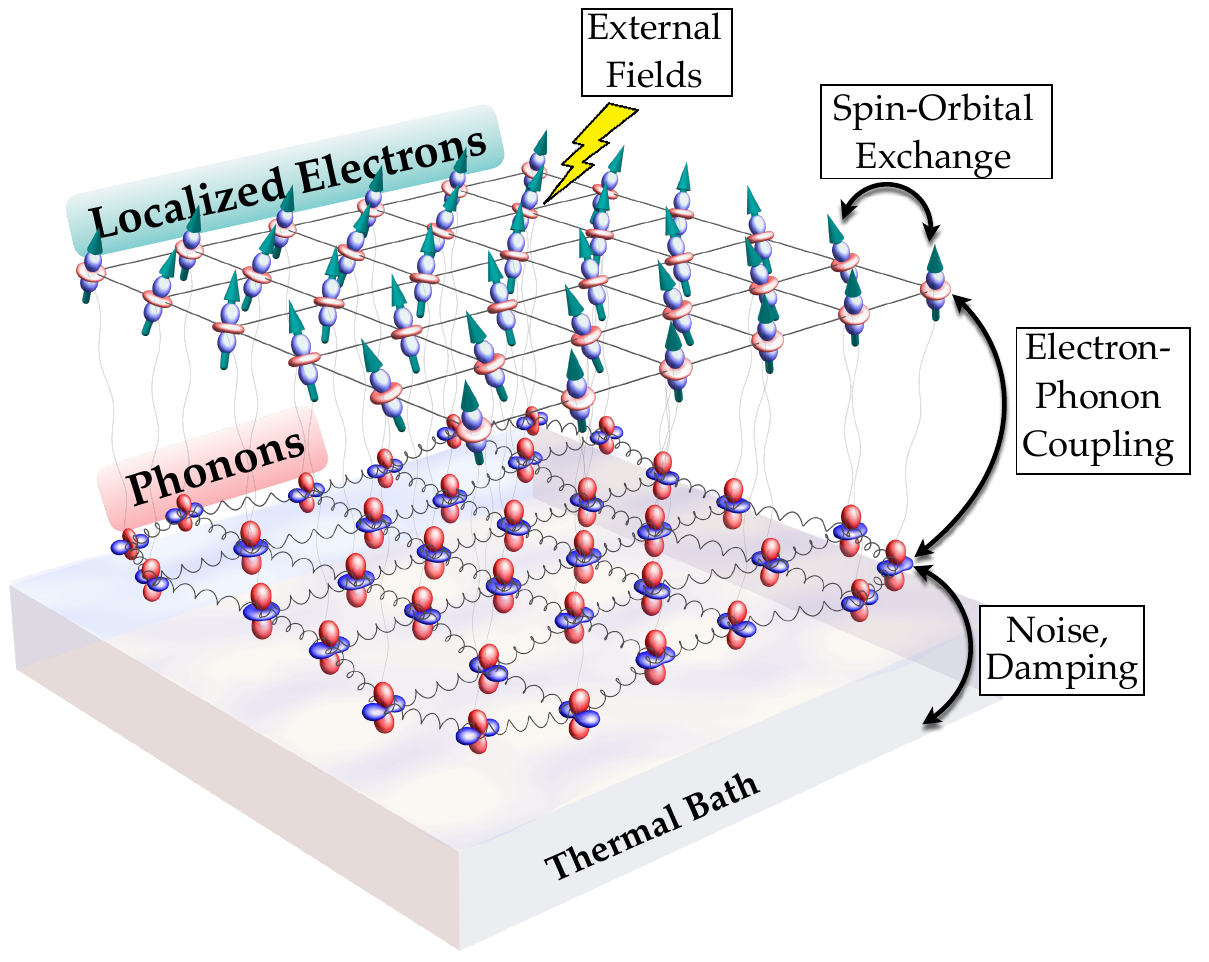} 	
    \caption{
    Schematic illustration of a spin–orbital Mott insulator 
    coupled to a thermal bath via phonon modes.
    Localized electrons interact via spin–orbital exchange 
    and experience local onsite terms, such as external fields 
    or single-ion anisotropies.
    Electron–phonon coupling mediates interactions with 
    lattice vibrations, which dissipate energy into a 
    thermal bath through damping and are subject to thermal noise.
    The illustration depicts a two-dimensional surface geometry for 
    visual clarity. 
    While the present framework is applicable to systems 
    of arbitrary dimensionality, the simplified bath coupling considered 
    here effectively describes phonon dissipation at the sample boundary.
    } 
\label{fig:concept}
\end{figure}

%
Describing the dynamical behavior of real materials remains a 
central challenge in condensed matter physics,  as it is essential for 
understanding, designing, and predicting materials 
functionality~\cite{Adler2019, Marzari2021, Cheng2025}.
Among quantum materials, Mott insulators provide a paradigmatic setting 
in which strong correlations 
suppress metallic behavior, leaving spin and orbital degrees of freedom to 
govern the low-energy physics.
This makes them a natural platform for correlation-driven phenomena 
such as magnetism~\cite{Fazekas-Book, FrustratedMagnetism-Book, Castelnovo2012}, 
spin liquids~\cite{Balents2010, Zhou2017, Savary2017, Knolle2019, Clark2021, SpinIce-Book}, 
and orbital order~\cite{Khomskii2003, Hotta2006, Streltsov2017, Khomskii2021}.

%
A systematic route to studying Mott insulators is provided by first-principles 
approaches, in which multiorbital Hubbard models are derived via downfolding 
to effective low-energy descriptions that capture the essential physics of 
the material~\cite{Marzari1997, Aryasetiawan1998, Miyake2009}.
Over the past decades, substantial progress has been made by combining 
first-principles calculations with effective model approaches, enabling a 
more direct connection between theoretical predictions and experimental 
observations of 
spin~\cite{Zhang12,Chiesa13,Yamaji14,Rau14,Winter16,Kurzydlowski17,Huang20,Fiore-Mosca22} and 
orbital~\cite{Pavarini08,Snamina16,Jeanneau17,Aligia19,Iwazaki2023} 
dynamics.
At the same time, significant efforts have been devoted to incorporating 
lattice vibrations, relaxation, and thermalization processes through electron–phonon 
interactions within first-principles frameworks~\cite{Giustino2017, Zhou2021}. 
However, such approaches remain computationally demanding and are currently 
limited in their ability to access long-time nonequilibrium dynamics in strongly 
correlated materials.
As a result, a fully microscopic description of dissipative electron 
dynamics—particularly far from equilibrium—remains an open challenge in 
materials research.

Meanwhile, stochastic Landau–Lifshitz–Gilbert (LLG)–type approaches 
provide a computationally efficient framework for simulating equilibrium and 
semiclassical dynamics of Mott 
insulators~\cite{Landau1992, Gilbert2004, Tsai2008}, 
including multiorbital systems with arbitrary 
spin $S$~\cite{Perelomov1972, Gitman1993, Nemoto2000, Zhang2021, Dahlbom2022a, Dahlbom2022b}, 
where the electronic degrees of freedom form an SU($N$) 
algebra with $N=2S+1$.
In suitable limits, such approaches can reproduce equilibrium properties in 
good agreement with quantum results after appropriate 
renormalization~\cite{Dahlbom2023, Kim2025}.

However, even within generalized SU($N$) formulations, energy 
dissipation is typically introduced 
phenomenologically~\cite{Gilbert2004, Dahlbom2022b}, 
without a direct microscopic foundation.
Recent developments combining SU($N$) stochastic Monte Carlo methods with 
mean-field dynamics~\cite{Hart25, Sutcliffe25} have highlighted 
the important role of phonons in driving 
spin and orbital dynamics in multiorbital Mott insulators.
Despite this progress, a fully microscopic framework that 
consistently captures electron–phonon coupling, dissipation, 
and relaxation in strongly correlated systems remains lacking.
Consequently, while classical and generalized SU($N$) approaches 
provide valuable insights, they remain limited in their ability to describe 
nonequilibrium dynamics and phonon-mediated dissipation on a microscopic 
footing in realistic Mott insulators.

In this paper, we develop electron–phonon coupled Langevin 
dynamics (epLD), a microscopic theory for describing the 
nonequilibrium dynamics of spin–orbital coupled strongly correlated
insulators.
By incorporating electron–phonon interactions, as schematically illustrated 
in Fig.~\ref{fig:concept}, our approach leads to generalized equations of motion.
Starting from an effective low-energy electronic 
Hamiltonian of the Kugel-Khomskii type~\cite{Kugel1972, Kugel1973, Iwazaki2023}, 
we formulate the dynamics 
using SU($N$) coherent states to enable the treatment of general spin-orbital 
degrees of freedom.
The derivation is carried out within the Keldysh path integral 
formalism~\cite{Kamenev_book,Mitra05,Picano23-1,Picano23-2}, which provides
a systematic route to capture both dissipative and stochastic effects arising from the 
coupling to phonons.
By taking the semiclassical limit of the quantum transition amplitude on 
the Keldysh contour, 
we obtain deterministic equations of motion augmented by a damping and a thermal 
noise terms--both derived microscopically from the underlying electron-phonon coupling.

We benchmark our formalism and its numerical implementation on a two-orbital 
spin chain coupled to Einstein phonons and demonstrate the hybridization of 
electronic and phononic bands as a function of electron-phonon coupling strength.
Finally, we show that our approach recovers the traditional LLG equations 
of motion in a specific limit.
Our framework enables simulations of both equilibrium and nonequilibrium 
dynamics in realistic strongly correlated materials, capturing dissipation, 
decoherence, and thermalization phenomena, providing a pathway to 
connect first-principles electronic band-structure calculations with 
microscopic dynamical behavior.

The remainder of this paper is organized as follows.
In Sec.~\ref{sec:Formalism}, we set up the path-integral formalism 
and derive the semiclassical equations of motion, 
which describe the general dynamics of an electron-phonon-coupled 
multiorbital Mott insulator.
These equations depend on the full time history and therefore 
describe a non-Markovian process, making them impractical to evaluate 
directly.
In Sec.~\ref{sec:NonMarkov.to.Markov}, we transform these non-Markovian 
equations of motion 
into a Markovian form 
by introducing auxiliary variables and considering the high-temperature 
limit of the statistical noise from the bath.
In this way, we obtain equations of motion [Eq.~\eqref{eq:eom.markov}], 
which represent the setup illustrated in Fig.~\ref{fig:concept} in a 
convenient form suitable for numerical simulations.
In Sec.~\ref{sec:Num.Benchmark}, we solve the equations of motion 
for a two-orbital ferromagnetic spin chain 
and benchmark the optimization process, thermodynamic observables, 
and dynamical structure factors against known results. 
We also explicitly demonstrate the hybridization between electronic 
and phonon bands as a function of electron-phonon coupling strength.
In Sec.~\ref{sec:relation.to.LLG}, we discuss the relation of our 
electron-phonon formalism to the widely used Landau-Lifshitz-Gilbert 
(LLG) equation. 
We show that the LLG equation emerges as a limiting case of our 
approach.
We conclude in Sec.~\ref{sec:Discussion} with a discussion of our 
results, their implications for the field, and possible extensions 
and future directions, followed by a summary in 
Sec.~\ref{sec:Summary}.

\section{Path Integral Formalism on the Keldysh Contour}
\label{sec:Formalism}

The goal of this work is to derive real-time equations of
motion for spin–orbital–coupled strongly correlated
insulators from a microscopic treatment of electron–phonon interactions.
To achieve this, we represent localized electrons as
SU($N$) coherent states
and solve their dynamics via a path integral 
formalism on the Keldysh contour.
As illustrated in Fig.~\ref{fig:concept}, we consider three 
coupled subsystems: \\
%
\begin{enumerate}
    \item localized electrons on a lattice, 
    interacting via spin-orbital exchange,
    \item phonon modes representing lattice vibrations, and
    \item a thermal bath at fixed temperature $T$.
\end{enumerate}
%
All three components are treated explicitly and are dynamically coupled.
Electrons and phonons interact through electron–phonon coupling, 
while the phonons themselves are coupled to a thermal bath. 
This phonon–bath interaction introduces both damping and stochastic noise, 
modeled microscopically after introducing auxiliary fields in the path integral 
formalism. 
These terms satisfy the fluctuation–dissipation theorem, ensuring that 
the phonons relax toward the correct thermal equilibrium distribution 
at temperature $T$.

The resulting equations of motion for electron–phonon coupled 
Langevin dynamics (epLD) take the form of two 
coupled first-order differential equations
(see Sec.~\ref{sec:NonMarkov.to.Markov} for details):
%
\begin{subequations}
    \begin{align}
        &\dot{\bm \Omega} = {\bm B}^{-1} \ \nabla_{\bm \Omega} {\bm{\mathcal{O}}}({\bm \Omega}) \ 
                \Big[ \underbrace{ 
                {\bm E}  
                - {\bm I} \ \bm{\mathcal{O}} 
                    ({\bm \Omega})}_{\text{electronic forces}}
                - \underbrace{g\big(\bm a 
                    + \bm a^*)}_{\text{ep coupling}}
                \Big]
                \, ,   
    \end{align}
    \begin{align}
        &\imu \Big( 1+\underbrace{\frac{\imu \gamma }{\omega}}_{\text{damping}} \Big)  \ \dot{\bm a} 
        = \omega {\bm a} 
                    + \underbrace{g \bm{\mathcal{O}}
                        ({\bm \Omega})}_{\text{ep coupling}} 
                    + \underbrace{\bm{\Gamma}}_{\text{noise}} \, .
    \end{align}
\end{subequations}
%
%
%
%
Here, $\bm{\Omega}$ parametrizes the electronic degrees of 
freedom via the spin--orbital mechanical variables
$\bm{\mathcal{O}}(\bm{\Omega})$, while $\bm a$ describes 
the local phonon mode. 
The electronic dynamics are governed by local terms $\bm E$,
interactions $\bm I$, and the Berry curvature matrix $\bm B$, 
and are coupled to the phonons through the electron–phonon 
interaction $g$.
The phonons, characterized by eigenfrequency $\omega$, 
undergo damped dynamics with rate 
$\gamma$ and are driven by thermal noise $\bm{\Gamma}$, 
reflecting their coupling to a thermal bath.
This structure provides a transparent picture of 
nonequilibrium energy flow: energy is transferred from 
the electronic system to the phonons and subsequently 
dissipated into the thermal environment.

\subsection{Hamiltonian}

In the following, we set up the generalized nonequilibriumal model 
used to describe the schematic concept shown 
in Fig.~\ref{fig:concept}.
The total Hamiltonian is written as 
%
\begin{equation}
    \hat{\mathscr H} = \hat{\mathscr H}_e + \hat{\mathscr H}_p + \hat{\mathscr H}_b
    + \hat{\mathscr H}_{ep} +\hat{\mathscr H}_{pb} 
    \, ,
\label{eq:Ham.all}
\end{equation}
%
where $\hat{\mathscr H}_e$, $\hat{\mathscr H}_p$, and 
$\hat{\mathscr H}_{b}$ represent the Hamiltonians of the 
electronic system, the phonons, and the thermal bath, respectively.
The term $\hat{\mathscr H}_{ep}$ 
describes the coupling between electrons and phonons, while 
$\hat{\mathscr H}_{pb}$ accounts for the coupling between 
phonons and the bath.

The electronic part is modeled as a nonequilibrium system, 
following the structure of the generalized Kugel-Khomskii 
Hamiltonian~\cite{Kugel1972, Kugel1973, Iwazaki2023}:
%
\begin{equation}
     \hat{\mathscr H}_e = \sum_{ ij  {\xi\xi'}} I_{ij}^{\xi\xi'} 
                    \hat{\mathscr O}_i^\xi \hat{\mathscr O}_j^{\xi'} 
                  - \sum_{i\xi} E_{i}^\xi \hat{\mathscr O}_i^\xi  \, ,
    \label{eq:e-ham}
\end{equation}
%
where $\hat{\mathscr O}_i^\xi$ denotes the operator 
of a local electronic state at site $i$, and 
$\xi$ labels internal quantum degrees of freedom 
(e.g., spin, orbital, or multipolar components),
whose precise definition will be given in 
Sec.~\ref{sec:coherent.states}.
The summation $\sum_{ij}$ runs over all interacting 
site pairs. 
The coupling tensor $I_{ij}^{\xi\xi'}$ encodes 
spin-orbital exchange interactions between electrons.
The single-site term ${E}_{i}^\xi$ accounts for local energy 
contributions, such as single-ion anisotropies or time-dependent 
external magnetic fields.

The phonon Hamiltonian is given by 
%
\begin{equation}
    \hat{\mathscr H}_p = \sum_{\bm{k} n} \omega_{\bm{k} n} \ \hat{a}^\dg_{\bm{k} n} \hat{a}_{\bm{k} n}  \, ,
    \label{eq:phonon-ham}
\end{equation}
%
where $\hat{a}^\dg_{\bm{k} n}$ and $\hat{a}_{\bm{k}}$ are the creation 
and annihilation operators for a phonon in mode $({\bm k}, n)$,
with eigenfrequency $\omega_{{\bm k} n}$.
Here, ${\bm k}$ is the phonon wavevector and $n$ is the phonon 
branch (or band) index.

The electron–phonon coupling takes the form
%
\begin{equation}
    \hat{\mathscr H}_{ep} = \sum_{i \xi} \sum_{\bm{k} n} \hat{\mathscr O}_i^{\xi} 
        \left( g_{i, \bm{k} n}^{\xi *} \hat{a}_{\bm{k} n} + \text{H.c.} \right) \, ,
    \label{eq:Ham_ep_coupling}
\end{equation}
%
which we refer to as \emph{site-phonon coupling},
where $g_{i, \bm{k} n}^{\xi}$ denotes the coupling 
strength between the local electronic operator $\hat{\mathscr O}_i^\xi$ and 
the phonon mode $(\bm{k}, n)$.
Here, we assume that the electron dynamics are much faster 
than the lattice dynamics (i.e., the Born–Oppenheimer-like separation 
of timescales), justifying a local, onsite form of the coupling 
without explicit intersite contributions 
in the electronic part.

Although not considered further in this work, the formalism can also
be generalized to \emph{bond-phonon coupling}, in which
Eq.~\eqref{eq:Ham_ep_coupling} is formulated in terms of electronic bond
operators 
$\hat {\mathscr H}_{ep}\sim\sum_{ij\bm k} g_{ij,\bm k} \hat c_i^\dg \hat c_j \hat a_{\bm k} + {\rm H.c.}$ 
where $\hat{c}_i^\dg, \hat{c}_i$, are the usual creation, annihilation operators of electrons.
For a magnetic spin-$1/2$ system, phonons cannot couple
directly to the local spin operator 
$\hat{\bm S}_i = \frac 1 2 \hat c_i^\dg \bm \sg \hat c_i $ with a Pauli matrix $\bm \sg$.
As a result, spin--phonon interactions arise only through
virtual charge fluctuations and are therefore mediated by
intersite processes.
Such processes arise from second-order perturbation theory involving virtual charge
excitations with an energy scale set by the onsite Coulomb
interaction $U$.
%
%
To leading order this yields the 
effective spin-phonon Hamiltonian:
\begin{equation}
    \hat{\mathscr H}_{ep}^{\rm eff} \sim 
        \frac{t_{ij}}{U} \ g_{ij, \bm{k}} \ \hat {\bm S}_i\cdot \hat{\bm S}_j
                (\hat{a}_{\bm{k}} + \hat{a}_{\bm{k}}^\dg) \, .
\label{eq:bond-phonon.energy}
\end{equation}
%
%
In contrast, for systems with spin $S \geq 1$ or with active 
orbital (or multipolar) degrees of freedom, direct onsite coupling 
to phonons --- as described in Eq.~\eqref{eq:Ham_ep_coupling} --- can 
become dominant. 
The corresponding local multipole moments
couple directly to lattice distortions without requiring charge 
fluctuations, naturally motivating
our focus on local electron–phonon coupling in this work.

We finally consider the coupling between the system’s phonons and 
those of an external thermal bath.
While a fully microscopic treatment would model phonon exchange 
across the system-bath interface, we adopt a simplified 
form of linear coupling
%
\begin{align}
    \hat{\mathscr H}_{b} &= \sum_\ell \omega_{\ell}' \hat{b}_{\ell}^\dg \hat{b}_{\ell}  \, , 
    \label{eq:bath.phonon-ham}
    \\
    \hat{\mathscr H}_{pb} &= \sum_{\ell }\sum_{\bm kn}  
                        (g'_{\ell,\bm kn} \hat{b}_{\ell}^\dg \hat{a}_{\bm kn} + {\rm H.c.} ) \, . 
    \label{eq:Ham_pb_coupling}
\end{align}
%
Here, $\hat{b}_{\ell}^\dg$ and $\hat{b}_{\ell}$ are the creation and 
annihilation operators for bath phonons with mode index
$\ell$ and frequency $\omega_{\ell}'$.
The coupling constant $g'_{\ell,\bm kn}$  characterizes 
the interaction between bath mode $\ell$ and system phonon 
mode $(\bm k, n)$, enabling energy exchange between the 
system and its thermal environment.
A more detailed discussion of realistic system-bath coupling mechanisms is presented in Sec.~\ref{sec:Discussion_reaslistic_bath}.

\subsection{SU($N$) Coherent States}
\label{sec:coherent.states}

In our formalism, electron interactions are 
described within a general multiorbital framework 
using an SU($N$) coherent state representation. 
This approach -- built on foundational work in 
Refs.~\cite{Perelomov1972, Nemoto2000, Zhang2021, Iwazaki2023} -- 
captures the full range of local electronic degrees of 
freedom, including higher-order multipolar 
moments.
The local operator $\hat{\mathscr O}_i^\xi$, as used in 
Eqs.~\eqref{eq:e-ham} and \eqref{eq:Ham_ep_coupling}, 
is defined as
%
\begin{equation}
    \hat{\mathscr O}_i^\xi = \sum_{\al,\beta=1}^N |\al\ra_i \ 
                        O^\xi_{\al\beta} \ {}_i\la \beta |  \, ,
\label{eq:O.coheren.operator}
\end{equation}
%
where $N$ is the dimension of the local Hilbert space. 
$|\alpha\rangle_i$ denotes a local basis state, 
and $O^\xi_{\alpha\beta}$ are the complex-valued matrix elements
of the SU($N$) generators in this basis, forming a complete set 
of traceless operators in the extended (Liouville) space.

The expectation value of the operator $\hat{\mathscr O}_i^\xi$ 
is given by
%
\begin{equation}
    \mathscr O_{i}^\xi (\Omega_i) = {}_i\la \Omega_i | 
                                    \hat{\mathscr O}_i^{\xi} |\Omega_i \ra_i \, ,
   \label{eq:O.coherent}
\end{equation}
%
with the SU($N$) coherent state
%
\begin{equation}
    |\Omega_i\ra_i = \sum_{\al=1}^N c_\al (\Omega_i) |\al\ra_i \, .
    \label{eq:def_of_c}
\end{equation}
%
The coefficients $c_\al (\Omega_i)$ parametrize the local 
quantum state.
While this representation is not unique, 
we adopt the parametrization introduced by
Nemoto~\cite{Nemoto2000} and Iwazaki {\it et al.} \cite{Iwazaki2023}
for general SU($N$) coherent states
%
\begin{align}
    c_{\alpha}(\Omega) = e^{i \varphi_{\alpha -1}} \cos{x}_{\alpha} 
                        \prod_{\beta = 1}^{\alpha-1}
                        \sin{x_{\beta}}       \, ,
\label{eq:c.alpha}
\end{align}
%
with the constraints $\varphi_0 = 0$ and $x_N = 0$, and parameter domains
%
\begin{align}
   x_1, &\cdots, x_{N-1} \in [0, \pi/2]  \, , 
   \label{eq:param.x}\\
   \varphi_1, &\cdots, \varphi_{N-1} \in [0, 2\pi)    \, .
   \label{eq:param.phi}
\end{align}
%
This construction ensures the normalization of the coherent state 
and provides a smooth, continuous parametrization of the SU($N$)
projective space~\cite{Iwazaki2023}.
Accordingly, this parametrization defines a set of real parameters
%
\begin{align}
     \Omega_i 
     &= 
     \{\Omega_{ip}\} \, , 
     \\
             &= (x_1,...,x_{N-1}, \varphi_{1}, \cdots, \varphi_{N-1} )_i  \, ,
\label{eq:Omega.set}
\end{align}
%
with $p=1,\cdots, 2(N-1)$ corresponding to the local electronic 
degrees of freedom [see Eqs.~\eqref{eq:param.x} and \eqref{eq:param.phi}]
at site $i$.
In the following, we omit the site index from bras and kets when the 
context 
is unambiguous, and write Eqs.~\eqref{eq:O.coherent} 
and \eqref{eq:def_of_c} 
simply as 
$\mathscr O^\xi_i (\Omega_i) = \mathscr O^\xi (\Omega_i)$ and 
$|\Omega_i\ra_i = |\Omega_i\ra$.

\subsection{Imaginary-Time Path Integral}

A common simplification is to treat phonons as 
classical variables with phenomenological damping.
However, this approach lacks a clear microscopic foundation.
We therefore go beyond this approximation by starting 
from a fully quantum-mechanical formulation and 
derive the equations of motion by taking the semiclassical 
limit at a later stage.

We begin by formulating the partition function as an 
imaginary-time path integral, treating phonons and 
electrons explicitly.
The total partition function is given as
%
\begin{align}
    Z &= \int 
    	  \mathscr D[\Omega, a, b]
       \, \epn^{- \mathscr S[\Omega, a, b] }  \, ,
    \label{eq:Z_def}
\end{align}
%
where the action $\mathscr S[\Omega, a, b]$ includes all contributions from 
the Hamiltonians defined in Eq.~\eqref{eq:Ham.all} and takes the 
explicit form
%
\begin{align}
    \mathscr S[\Omega,a,b] &= \int_0^\beta \diff\tau \Big[\sum_{i\al} c^*_\al(\Omega_i)\partial_\tau c_\al (\Omega_i)
    \nonumber \\
    &\ \ \ 
    + \sum_{\la ij \ra{\xi\xi'}} I_{ij}^{\xi\xi'} \mathscr O^\xi (\Omega_i) \mathscr O^{\xi'}(\Omega_j) - \sum_{i\xi} E_{i}^\xi \mathscr O^\xi (\Omega_i)
    \nonumber \\
    &\ \ \ 
    + \sum_{i\xi}\sum_{\bm kn} \mathscr O^\xi(\Omega_i) \qty( g_{i,\bm kn}^{\xi*} a_{\bm kn}+ {\rm H.c.}  )
    \nonumber \\
    &\ \ \  + \sum_{\bm kn} \qty(
    a_{\bm kn}^* \partial_\tau a_{\bm kn}
    + \omega_{\bm kn} a_{\bm kn}^* a_{\bm kn}
    ) 
    \nonumber \\
    &\ \ \ 
    + \sum_{\ell}\sum_{\bm kn}  \qty( g'_{\ell,\bm kn}b_{\ell}^* a_{\bm kn}+ {\rm H.c.}  )
    \nonumber \\
    &\ \ \  + \sum_{\ell} \qty(
    b_{\ell}^* \partial_\tau b_{\ell}
    + \omega'_{\ell} b_{\ell}^* b_{\ell}
    ) \Big]  \, ,
    \label{eq:action_orig}
\end{align}
%
with $\beta = 1/T$ the inverse temperature.
All imaginary-time dependence enters through the real-valued SU($N$) 
coherent state parameters $\Omega_i(\tau)$ 
[see Eqs.~\eqref{eq:def_of_c}--\eqref{eq:Omega.set}] and through the 
complex-valued fields $a_{\bm kn}(\tau)$ and $b_{\ell}(\tau)$, 
which correspond to the 
eigenvalues of the phonon operators in Eqs.~\eqref{eq:phonon-ham} and 
\eqref{eq:bath.phonon-ham}, respectively. 
The time-derivative terms $c^*_\al(\Omega_i)\partial_\tau c_\al (\Omega_i)$,
$a_{\bm kn}^* \partial_\tau a_{\bm kn}$, and 
$b_{\ell}^* \partial_\tau b_{\ell}$
arise from quantum dynamics and reflect the geometric nature of 
the path in phase space.

The phonon fields appear in both linear and bilinear forms in 
the action, allowing them to be integrated 
out exactly using Gaussian path integration after a Fourier transform.
We begin by integrating out the bath phonons $b_{\ell}$, which yields 
the following contribution to the action
%
\begin{align}
    \sum_\nu \sum_{\bm kn}\sum_{\bm k'n'} 
    a_{\bm kn}^* (\imu\nu) \ 
    \Pi_{\bm kn,\bm k'n'}(\imu\nu) \ 
    a_{\bm k'n'}(\imu\nu)        \, ,
    \label{eq:Pi_self_energy}
\end{align}
%
where the Fourier transformation for the bosonic field is defined by
%
\begin{align}
    a_{\bm kn}(\imu\nu) = \frac{1}{\sqrt \beta} \int_0^\beta \diff \tau \  a_{\bm kn}(\tau) \epn^{\imu \nu \tau} \, ,
     \label{eq:Fourier_def}
\end{align}
%
with $\nu$ being a bosonic Matsubara frequency.
The resulting 
system-phonon self-energy originating from coupling to the bath
is given by
%
\begin{align}
    \Pi_{\bm kn,\bm k'n'}(\imu\nu)
    &=
    \sum_{\ell}
    \frac{{g'}^{*}_{\ell,\bm kn} g'_{\ell,\bm k'n'}}{\imu \nu -\omega_{\ell}}     \nonumber \\
    &\approx  \frac{\gm_{\bm kn}}{\omega_{\bm kn}} |\nu| \delta_{\bm k\bm k'} \delta_{nn'}  \, ,
    \label{eq:Pi_simplified}
\end{align}
%
where we assume a simplified, local damping model for 
the bath, as the microscopic structure of the coupling $g'_{\ell,\bm kn}$
is generally unknown. 
Within this approximation, the bath behaves as an Ohmic 
environment~\cite{Kamenev_book}, such that each phonon mode is damped 
independently with a damping rate $\gm_{\bm kn}$.
Note that a constant (real-valued) term may also appear in 
Eq.~\eqref{eq:Pi_simplified}, which, however, can 
be absorbed into a redefinition 
of the phonon single-particle energy and therefore does not play 
an explicit role in our formalism.

We now proceed to integrate out the system phonons $a_{\bm kn}$,
leading to an effective action for the electronic degrees of freedom
%
\begin{align}
    \mathscr S_{\rm eff} &= \mathscr S_0 + \mathscr S_{\rm diss}  \, ,     
    \label{eq:Action.Effective}
\end{align}
%
where the first term describes the isolated electron system 
%
\begin{align}
    \mathscr S_0 &= \int_0^\beta \diff\tau \left[ \sum_{i\al}  c^*_\al(\Omega_i)\partial_\tau c_\al (\Omega_i)
					+ \mathscr H_e(\bm \Omega) \right] \, .
\label{eq:action.matrsubara.unperturbed}
\end{align}
%
Here, \mbox{$\bm \Omega = \{ \Omega_i \}$} denotes the set of coherent-state 
parameters across all lattice sites.
The term $\mathscr H_e(\bm \Omega)$ is the energy functional corresponding 
to the electronic Hamiltonian [second line of Eq.~\eqref{eq:action_orig}].
The second term in Eq.~\eqref{eq:Action.Effective}
encodes the dissipative effects arising from the phonons and takes the form
%
\begin{align}
    &\mathscr S_{\rm diss}
    = \sum_\nu \sum_{ij\xi\xi'}
     \mathscr O_i^{\xi*}(\imu\nu)\Sigma_{i\xi,j\xi'} (\imu \nu) 
      \mathscr O_j^{\xi'} (\imu\nu)	\, .
\label{eq:S.diss.Matsubara}
\end{align}
%
Here, the system-phonon self-energy $\Sigma$ is given by 
%
\begin{align}
    \hspace{-2mm} \Sigma_{i\xi,j\xi'}(\imu \nu) 
     =
     \frac 1 2 \sum_{\bm kn} \Big[ g_{i,\bm kn}^{\xi* } g_{j,\bm kn}^{\xi'} G_{\bm kn} (\imu\nu)
     + g_{i,\bm kn}^{\xi } g_{j,\bm kn}^{\xi'*} G_{\bm kn} (-\imu\nu) \Big] \, ,
\end{align}
%
where the dressed phonon Green's function is
%
\begin{align}
	G_{\bm kn}(\imu \nu) 
		&=	\frac{1}{\imu\nu - \omega_{\bm kn} - \gamma_{\bm kn} |\nu| /\omega_{\bm kn}} \, .
\label{eq:phonon.green}
\end{align}
%
This result captures both the retarded propagation of phonons and 
the dissipative influence of the thermal bath, encoded through the 
damping rate $\gm_{\bm kn}$.
Physically, this damping term reflects the energy transfer from the system 
phonons to the external environment.
While a more detailed treatment would require specifying how the system 
couples to a physical reservoir (e.g., a substrate or cryostat), we 
adopt this simplified, mode-resolved damping description 
[see Eq.~\eqref{eq:Pi_simplified}]
as a practical and commonly used approximation.
For an idealized system, the damping can be turned off by taking 
the limit $\gm_{\bm kn} \to 0$.

Using the phonon Green’s function from Eq.~\eqref{eq:phonon.green}, 
we now express the dissipative term from Eq.~\eqref{eq:S.diss.Matsubara}
in imaginary time and rewrite Eq.~\eqref{eq:Action.Effective} 
as 
%
\begin{align}
    \mathscr S_{\rm eff} &= \mathscr S_0   \nonumber \\
    &+ \int\diff \tau \diff \tau' \sum_{ij\xi\xi'}
    \mathscr O^{\xi}_i(\tau) \Sigma_{i\xi,j\xi'}(\tau-\tau') \mathscr O^{\xi'}_j (\tau') \, ,
\label{eq:gen_form_lagrangian}
\end{align}
%
with time dependence expressed through the 
inverse Fourier transform
[c.f. Eq.~\eqref{eq:Fourier_def}].
The time dependence in the self-energy, $\Sigma(\tau-\tau')$, captures 
memory effects and dissipation 
resulting from the coupling of electrons to system phonons and the bath.

\subsection{Real-Time Keldysh Formalism}
\label{sec:real_time_keldysh}

We now consider the real-frequency representation of the self-energy
in Eq.~\eqref{eq:gen_form_lagrangian}.
The retarded self-energy is obtained by analytic continuation of the 
Matsubara self-energy via $\imu \nu \to \ep + \imu 0^+$, for $\nu>0$,
resulting in 
%
\begin{align}
   &  \Sigma^{\rm R}_{i\xi,j\xi'} (\ep)  
    = 
    \frac 1 2 \sum_{\bm kn} \Big[ g_{i,\bm kn}^{\xi *} g_{j,\bm kn}^{\xi'} G^{\rm R}_{\bm kn} (\ep)
     + g_{i,\bm kn}^{\xi } g_{j,\bm kn}^{\xi'*} G^{\rm A}_{\bm kn} (-\ep) \Big]  
     \, .
\label{eq:retarded_expr2}
\end{align}
%
The advanced self-energy follows from $\imu \nu \to \ep - \imu 0^+$ 
for $\nu<0$, giving 
%
\begin{align}
    \Sigma_{i\xi,j\xi'}^{\rm A}(\ep) = 
    \Sigma_{j\xi',i\xi}^{\rm R}(\ep)^*  \, .
\label{eq:self-energy.retarded.advanced}
\end{align}
%

%
Its corresponding real-time representation is obtained via Fourier 
transform of the retarded and advanced Green's functions
%
\begin{align}
    G^{\rm R}_{\bm kn}(t)  =  G_{\bm kn}^{\rm A}(- t)^* 
    = - \frac{\imu\theta(t)}{z_{\bm kn}}
    \epn^{-\imu\omega_{\bm kn}t/z_{\bm kn}} \, ,
    \label{eq:GR_concrete_expr}
\end{align}
where we defined the dimensionless complex number
%
\begin{align}
	z_{\bm kn} \equiv 1 + \frac{\imu \gm_{\bm kn}}{\omega_{\bm kn}} \, .
    \label{eq:def_z}
\end{align}
The Heaviside step function $\theta(t)$, defined as 
$\theta(t) = 1$ for $t > 0$ and $\theta(t) = 0$ for $t < 0$,
ensures causality, meaning that the future evolution depends  
only on the past.
In the weak damping limit $\gm_{\bm kn} \ll\omega_{\bm kn}$, 
Eq.~\eqref{eq:GR_concrete_expr} 
reduces to the more familiar form
%
\begin{align}
    G_{\bm kn}^{\rm R}(t)\simeq -
    		\imu \theta(t) \epn^{-\imu\omega_{\bm kn}t - \gm_{\bm kn}t} \, ,
\label{eq:Greens.func}
\end{align}
confirming that 
$\gm_{\bm kn}$ indeed represents a damping rate.

In thermal equilibrium, the Keldysh component of the self-energy 
satisfies the fluctuation-dissipation relation~\cite{Kamenev_book}
%
\begin{align}
    \Sigma_{i\xi,j\xi'}^{\rm K} (\ep) &=
    \qty[ \Sigma_{i\xi,j\xi'}^{\rm R} (\ep) - \Sigma_{i\xi,j\xi'}^{\rm A} (\ep) ] \coth \frac{\ep}{2T} \, .
    \label{eq:Keldysh_def}
\end{align}
%
We note that this relation relies on the assumption that the 
system phonons remain close to thermal equilibrium. 
The case in which system phonons are far out of 
equilibrium will be discussed in Sec.~\ref{sec:Discussion_reaslistic_bath}.

Equation \eqref{eq:Keldysh_def}
can be expressed
compactly as
%
\begin{align}
    \Sigma_{i\xi,j\xi'}^{\rm K} (\ep)
    &= - \imu \sum_{\bm kn} \Big[ g_{i,\bm kn}^{\xi*} g_{j,\bm kn}^{\xi'} F_{\bm kn}(\ep)
    + g_{i,\bm kn}^{\xi} g_{j,\bm kn}^{\xi'*} F_{\bm kn} (-\ep) \Big]	\, ,
    \label{eq:Keldysh_expression}
\end{align}
%
with the spectral function $F_{\bm kn}(\ep)$ defined as
%
\begin{align}
	F_{\bm kn}(\ep) &= - \, {\rm Im\,} G_{\bm kn}^{\rm R} (\ep) \, \coth\frac{\ep}{2T}	\, .
\label{eq:Ffunc}
\end{align}
%

We now transform the effective action of Eq.~\eqref{eq:gen_form_lagrangian}
into its real-time representation. 
In this form, the action consists again of two parts
%
\begin{align}
    S_{\rm eff} = S_0 + S_{\rm diss}  \, .    
\end{align}
%
where $S_0$ is the isolated electronic system [c.f. Eq.~\eqref{eq:action.matrsubara.unperturbed}], 
and $ S_{\rm diss}$ captures 
dissipation induced by coupling to phonons and the thermal bath 
[c.f. Eq.~\eqref{eq:S.diss.Matsubara}].
%

%
\begin{figure}[tb]
    \centering
    \includegraphics[width=0.4\textwidth]{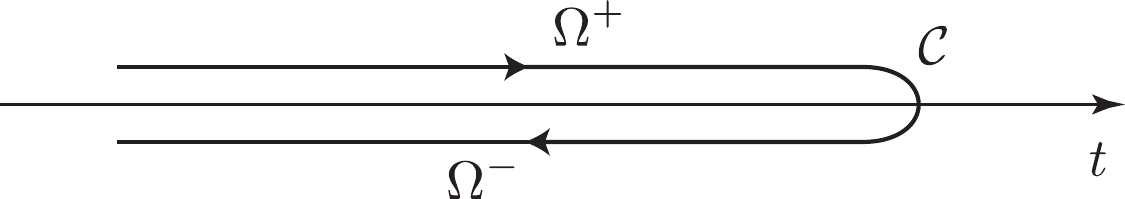} 	
    \caption{
    Schematic illustration of the Kelydysh contour 
    $\mathcal C$ as used in the integral of 
    Eq.~\eqref{eq:S.diss.Keldysh}.
    The contour extends from $-\infty$ to $+\infty$ 
    along both the forward and backward branches. 
    }
\label{fig:keldysh}
\end{figure}

The dissipative part is written as an interaction 
that is nonlocal in time along the Keldysh contour $\mathcal C$ 
%
\begin{align}
	S_{\rm diss}  &= \int_{\mathcal C}\diff t \diff t' \sum_{ij\xi\xi'}
			\mathscr O^{\xi}_i(t) \tilde \Sigma_{i\xi,j\xi'}(t,t') \mathscr O^{\xi'}_j (t') \, ,
\label{eq:S.diss.Keldysh}
\end{align}
%
where $\tilde \Sigma$ is the contour-ordered self-energy kernel.
The coherent-state fields are defined along the forward and backward branches 
of the contour 
$\Omega_i^+$ for $(-\infty, \cdots, +\infty)$ and 
$\Omega_i^-$ for $(+\infty, \cdots, -\infty)$ as  shown in Fig.~\ref{fig:keldysh}.

Applying the standard Keldysh rotation~\cite{Kamenev_book}, we define 
{\it classical} $(\rm cl)$ and {\it quantum mechanical} $(\rm qm)$ components of the  
coherent state variables and local electronic observables
%
\begin{align}
    \Omega_i^{\rm cl} &= 
    \left( \Omega_i^+ + \Omega_i^- \right) /2 \, ,
    \\
    \Omega_i^{\rm qm} &= \Omega_i^+ - \Omega_i^-  \, ,
    \label{eq:def_qm}
    \\
    \mathscr O_i^{\xi, {\rm cl}} &=  \qty[ \mathscr O^{\rm \xi} (\Omega_i^+) 
    					       + \mathscr O^{\rm \xi} (\Omega_i^-) ] /2 \, ,
    \label{eq:defO_cl}  \\
    \mathscr O_i^{\xi, {\rm qm}} &=  \mathscr O^{\rm \xi} (\Omega_i^+) 
    						- \mathscr O^{\rm \xi} (\Omega_i^-) \, .
    \label{eq:defO_qm}  
\end{align}
%
That is, the classical components are defined as the average of 
the forward and backward fields on the Keldysh contour, 
while the quantum components correspond to their difference.

Using this rotation and the Matsubara-to-Keldysh 
correspondence~\cite{Kamenev_book}, we rewrite both the system action 
and the dissipative term. 
The effective actions for isolated and dissipative parts
become
%
\begin{widetext}
\begin{align}
	S_0 &= \imu \int_{-\infty}^{\infty} \hspace{-2mm} \diff t 
            \sum_{s=\pm} \left[ s\sum_{i\al} c_\al^*
    	\qty( \Omega_i^{\rm cl}+ \tfrac 1 2 s\Omega_i^{\rm qm}) 
            \partial_t c_\al \qty(\Omega_i^{\rm cl}+ \tfrac 1 2 s\Omega_i^{\rm qm})
    	+s \imu \ {\mathscr H}_e \qty(
        \bm \Omega^{\rm cl}
            + \tfrac 1 2 s \bm \Omega^{\rm qm}
            )	\right]	\, ,
        \label{eq:S0.Keldysh.2}  \\
	\nonumber \\
	S_{\rm diss} &= - \sum_{ij\xi\xi'} \int_{-\infty}^{\infty} \diff t \ \diff t' 
  		  \begin{pmatrix}
       			 \sqrt 2\mathscr O^{\xi, {\rm cl} }_i & 
        			\mathscr O^{\xi, {\rm qm} }_i/\sqrt 2
   		 \end{pmatrix}_t
   		 \begin{pmatrix}
        			0 & \Sigma^{\rm A}_{i\xi,j\xi'} \\[2mm]
      			  \Sigma^{\rm R}_{i\xi,j\xi'} & \Sigma^{\rm K}_{i\xi,j\xi'}
   		 \end{pmatrix}_{tt'}
  		  \begin{pmatrix}
     			   \sqrt 2\mathscr O^{\xi', {\rm cl} }_j \\
     			   \mathscr O^{\xi', {\rm qm} }_j /\sqrt 2
   		 \end{pmatrix}_{t'} \, .
         \label{eq:S.diss.Keldysh.2}
\end{align}
\end{widetext}
%
This Keldysh-rotated form of the action makes causality manifest: the 
retarded and advanced components encode dissipative response, while 
the Keldysh component governs fluctuations~\cite{Kamenev_book}.
Importantly, this representation remains fully quantum mechanical, 
with no classical or semiclassical approximations invoked at this stage.

\subsection{Equations of Motion in the Semiclassical Limit}

We now focus on the dominant contributions to the path integral 
in the semiclassical limit.
The real-time dynamics are governed by the 
partition function 
%
\begin{align}
    Z=\int \mathscr D [\Omega] \, \epn^{\imu \hbar^{-1}S_{\rm eff}} \, ,
\end{align}
%
which plays the role of the time-evolution 
operator in the Keldysh path-integral formalism.
We rescale the quantum component in the functional 
integral as
$\Omega^{\rm qm} \to \hbar \Omega^{\rm qm}$.
This makes the expansion in $\hbar$ 
equivalent to an expansion in $\Omega^{\rm qm}$,
as discribed in Ref.~\cite{Kamenev_book}.
We then expand the classical and quantum mechanical
observables for electrons as
%
\begin{align}
    \mathscr O_i^{\xi, \rm cl} &= \mathscr O^{\xi} (\Omega_i^{\rm cl}) 
                    + O\qty( (\Omega^{\rm qm})^{1} )  \, ,   
    \label{eq:expand.cl} \\
    \mathscr O_i^{\xi, \rm qm} &= \sum_p \frac{\partial \mathscr O^{\xi} (\Omega_i^{\rm cl})}{\partial \Omega_{ip}^{\rm cl}} \Omega_{ip}^{\rm qm} 
                    + O\qty( (\Omega^{\rm qm})^{2} )  \, ,
    \label{eq:expand.qm}
\end{align}
%
where the sum over $p$ runs over the $2(N-1)$ 
real local electronic degrees of freedom parametrizing $\Omega_i$ 
[see Eq.~\eqref{eq:Omega.set}].
Substituting the leading-order terms 
from Eqs.~\eqref{eq:expand.cl} and \eqref{eq:expand.qm} 
into $S_0$ [Eq.~\eqref{eq:S0.Keldysh.2}], we obtain the 
semiclassical form of the action for 
the isolated electronic system
%
\begin{align} 
    S_0 \simeq \int \diff t \sum_{i p} \bigg( \sum_q \ \imu  B_{pq}(\Omega_i^{\rm cl})  \partial_t \Omega_{iq}^{\rm cl}
    - \sum_i \frac{\partial {\mathscr H}_e (\bm \Omega^{\rm cl})}{\partial \Omega_{ip}^{\rm cl}} \bigg) \Omega^{\rm qm}_{ip} \, ,
    \label{eq:S0_with_real_time}
\end{align}
%
with the Berry curvature matrix defined as
%
\begin{align} 
    B_{pq}(\Omega_i) = \sum_\al \qty[ \frac{\partial c^*_\al(\Omega_{i})}{\partial \Omega_{ip}} \frac{\partial c_\al(\Omega_{i})}{\partial \Omega_{iq}} - \frac{\partial c^*_\al(\Omega_{i})}{\partial \Omega_{iq}}\frac{\partial c_\al(\Omega_{i})}{\partial \Omega_{ip}} ] \, .
    \label{eq:def_Berry_curvature}
\end{align}
%

We divide the dissipative part of the action, 
$S_{\rm diss}$ [Eq.~\eqref{eq:S.diss.Keldysh.2}], into 
its retarded/advanced and Keldysh components
%
\begin{align} 
    S_{\rm diss} = S^{\rm R/A}_{\rm diss} 
                 + S^{\rm K}_{\rm diss} \, .
\end{align}
%
%
The first term, $S^{\rm R/A}_{\rm diss}$, captures 
dissipation and involves the real part of the retarded 
self-energy
%
\begin{align}
    S^{\rm R/A}_{\rm diss} =&
    2 \sum_{ij\xi\xi'} \int \diff t \ \diff t'
    \mathscr O_{i}^{\xi,{\rm qm}}(t)
   \ {\rm Re}\big[ \Sigma^{\rm R}_{i\xi,j\xi'} (t,t') \big] \ 
    \mathscr O_j^{\xi',{\rm cl}} (t') \, ,
\end{align}
%
where we used the complex-conjugate relation 
between $\Sigma^{\rm R}$ and $\Sigma^{\rm A}$ 
[see Eq.~\eqref{eq:self-energy.retarded.advanced}].
The quadratic terms in $\Omega^{\rm qm}$ 
give rise to  
thermal noise, captured by 
the Keldysh part of the action.
The Keldysh component is given by
%
\begin{align}
    \imu S_{\rm diss}^{\rm K} &= -
    \sum_{\bm kn} \int \diff \ep
    \ F_{\bm k n}(\ep)
    |V^{\rm qm}_{\bm kn}(\ep)|^2        \, ,
    \label{eq:SH_transforming_SK}
    \\
     V_{\bm kn}^{\al}(\ep) &= \frac{1}{\sqrt{2\pi}} \int \diff t \ 
        \epn^{\imu \ep t} \sum_{i\xi} g_{i,\bm kn}^\xi 
        \mathscr O_i^{\xi,\al}(t)       \, ,
    \label{eq:def_of_V}
\end{align}
%
with $\al = \rm \{qm, cl\}$, and 
$F_{\bm kn}(\ep)$ 
defined in Eq.~\eqref{eq:Ffunc}.

To linearize the Keldysh contribution
in $\Omega^{\rm qm}$, we perform a
Stratonovich-Hubbard (SH) transformation,
%
\begin{align}
    \epn^{\imu S_{\rm diss}^{\rm K}} &= \int \mathscr D [\zeta] \exp \qty[ - \sum_{\bm kn} \int \diff \ep \  \frac{|\zeta_{\bm kn}(\ep)|^2}{F_{\bm kn}(\ep)} ]
    \epn^{ \imu \tilde S_{\rm diss}^{\rm K}[\zeta]} \, ,
    \\
    \tilde S_{\rm diss}^{\rm K} &=
    \sum_{\bm kn} \int \diff \ep \big[
    \zeta_{\bm kn}^*(\ep)V^{\rm qm}_{\bm kn}(\ep) + \zeta_{\bm kn}(\ep)V^{{\rm qm}*}_{\bm kn}(\ep) \big] \, .
\end{align}
%
Here, $\zeta_{\bm kn}(\ep)$ is an auxiliary 
field that represents the stochastic noise spectrum.
For the Gaussian integral to converge, it is essential that
$F_{\bm kn}(\ep) > 0$, 
which can be verified using  
Eqs.~\eqref{eq:phonon.green} and \eqref{eq:Ffunc}.

We define the Fourier transformation of the SH field from 
frequency to time
%
\begin{align}
    \zeta_{\bm kn} (t) 
    &= \frac{1}{\sqrt{2\pi}} \int \diff \ep \ 
        \zeta_{\bm kn}(\ep) \epn^{-\imu \ep t } \, .
\end{align}
%
With this, the Keldysh action becomes
%
\begin{align}
    \tilde S_{\rm diss}^{\rm K} &= \sum_{i\xi}\int \diff t \ \mathscr O_i^{\xi ,\rm qm}(t) \sum_{\bm kn} \qty[ g_{i,\bm kn}^\xi \zeta^*_{\bm kn}(t) + {\rm c.c.}] \, .
\end{align}
%
We define the average as
%
\begin{align}
    \la \cdots \ra &= \frac{\int \mathscr D [\zeta]\  (\cdots) \exp \qty[ - \sum_{\bm kn} \int \diff \ep \frac{|\zeta_{\bm kn}(\ep)|^2}{F_{\bm kn}(\ep)} ]}{\int \mathscr D [\zeta]\  \exp \qty[ - \sum_{\bm kn} \int \diff \ep \frac{|\zeta_{\bm kn}(\ep)|^2}{F_{\bm kn}(\ep)} ]} \, ,
\end{align}
%
and obtain the noise spectrum
%
\begin{align}
    \la \zeta_{\bm kn}^*(\ep) \zeta_{\bm kn}(\ep') \ra 
    &=  F_{\bm kn} (\ep) \delta(\ep - \ep')   \, .
\label{eq:color_spectrum}
\end{align}
%

Let us now combine all components to derive the 
semiclassical equation of motion.
The partition function takes the form
%
\begin{widetext}
\begin{align}
    Z &= 
    \int \mathscr D [\zeta] \exp \qty[ - \sum_{\bm kn} \int \diff \ep \frac{ |\zeta_{\bm kn}(\ep)|^2}{F_{\bm kn}(\ep)} ]
     \int \mathscr D[\Omega^{\rm cl}, \Omega^{\rm qm}] \
    \exp \Bigg[
    \imu \sum_{ip} \int \diff t \ \Omega^{\rm qm}_{ip} 
    \ \mathcal{X}_{ip}(t)
        \Bigg] \, ,
    \label{eq:part_func_semiclas}   \\
    \mathcal{X}_{ip}(t) &=
    \sum_q \imu B_{pq}(\Omega_i^{\rm cl}) \dot \Omega_{iq}^{\rm cl} 
    - \frac{\partial {\mathscr H}_e (\bm \Omega^{\rm cl})}{\partial \Omega_{ip}^{\rm cl}} 
    + \sum_{\xi }
    \frac{\partial \mathscr O^{\xi,{\rm cl}}_i}{\partial \Omega_{ip}^{\rm cl} }
    \qty{
    2 \sum_{j\xi'}
    \int \diff t'
    \ {\rm Re} \left[ \Sigma^{\rm R}_{i\xi,j\xi'} (t,t') \right] \ 
    \mathscr O_j^{\xi',{\rm cl}} (t')
    + \sum_{\bm kn}\qty( g_{i,\bm kn}^\xi \zeta^*_{\bm kn}(t) + {\rm c.c.}) 
    } \, .
\end{align}
%

This form of the partition function reflects an underlying 
mathematical structure, similar to the well-known delta function 
identity
$\int_{-\infty}^{\infty} \diff q \, \epn^{\imu q x} = 2\pi \delta(x)$,
where only the contribution at $x=0$ survives.
Analogously, in our case, the functional integral
$\int \mathscr D[\Omega^{\rm qm}] 
\epn^{\imu \Omega^{\rm qm} \mathcal{X}}$,
is most significant for a configuration where 
$\mathcal{X}_{ip}(t) = 0$.
This condition defines the dominant semiclassical 
path within the partition function. 
Applying this reasoning, the semiclassical 
electron–phonon coupled Langevin dynamics (epLD) 
equation of motion becomes
%
\begin{align}
-\imu  \sum_{q} B_{pq}(\Omega_i) \dot \Omega_{iq} 
    & = \sum_{\xi} 
    \frac{\partial \mathscr O^{\xi}_i}{\partial \Omega_{ip} }  
    \qty[ 
    E_i^\xi 
    - 
    \sum_{j\xi'}
    \bigg(
    I_{ij}^{\xi\xi'} \mathscr O_j^{\xi'}
    +
    2 \int_{-\infty}^{t} \diff t'
    \ 
     \ {\rm Re}\left[ \Sigma^{\rm R}_{i\xi,j\xi'} (t-t') \right] \ 
    \mathscr O_j^{\xi'} (t')
    \bigg) 
    - \sum_{\bm kn} \qty(g_{i,\bm kn}^\xi \zeta_{\bm kn}^* + {\rm c.c.}) 
    ] \, ,
    \label{eq:eom}
\end{align}
\end{widetext}
where we have omitted the symbol ``cl'' 
for brevity and display only the $t'$-dependence 
explicitly.
The single-site energy term $E_i^\xi$ 
may, in general, be time-dependent, which allows the 
treatment of localized electron dynamics in 
the presence of time-dependent external fields.
The retarded self-energy $\Sigma^{\rm R}$ encodes both 
phonon-mediated interactions and dissipation 
effects.
The stochastic term involving $\zeta_{\bm kn}(t)$ 
corresponds to thermal noise, with statistics 
given by Eq.~\eqref{eq:color_spectrum}.

Note that the self-energy $\Sigma^{\rm R} (t-t')$ in Eq.~\eqref{eq:eom} 
introduces memory effects, making the dynamics inherently 
non-Markovian.
This nonlocality in time poses significant challenges for numerical simulation.
Fortunately, the non-Markovian problem can be mapped onto an 
equivalent Markovian process, as we will outline in detail in 
the next section.

Finally, we remark that our approach recovers known results in
appropriate limits.
In particular, for coherent states of SU(2), 
the semiclassical 
dynamics reduce to the standard stochastic LLG 
equation under specific conditions, as shall be demonstrated in detail in Sec.~\ref{sec:LLG.comparison}.
%

\section{From Non-Markov to Markov}
\label{sec:NonMarkov.to.Markov}

In actual numerical calculations, Markov processes---whose 
evolution depends only on the current state and not on the 
full history---are much easier to handle than history-dependent 
non-Markovian processes.
However, the  
noise spectrum in Eq.~\eqref{eq:color_spectrum} 
and the retarded self-energy in Eq.~\eqref{eq:eom} 
introduce memory effects, resulting in a non-Markovian process.
For certain classes of stochastic processes (a well-known 
example is the Ornstein-Uhlenbeck process~\cite{Risken_book}), 
it is established that non-Markovian dynamics can be mapped 
onto an equivalent Markovian representation by introducing 
auxiliary variables~\cite{Marchesoni83, Luczka06, Ceriotti09}.
A similar strategy can be applied to our equations of motion, 
enabling a reformulation in which auxiliary semiclassical 
degrees of freedom 
absorb the history-dependent memory effects.

\subsection{Retarded Interaction}

We construct a Markov-type representation 
for the retarded contribution 
$\Sigma^{\rm R}$ in the equation of motion \eqref{eq:eom}
by rewriting it as
%
\begin{align}
    2\sum_{j\xi'} \int_{-\infty}^t \hspace{-2mm} &\diff t' 
     \ {\rm Re}\left[ \Sigma^{\rm R}_{i\xi,j\xi'} (t-t') \right] \ 
    \mathscr O_j^{\xi'}(t')
    \nonumber \\
    = &\sum_{\bm kn} 
    g_{i,\bm kn}^{\xi *} 
    A_{\bm kn}(t) +{\rm c.c.}    \, ,
    \label{eq:introduction_of_A}
\end{align}
%
where we have introduced the auxiliary variable
%
\begin{align}
    A_{\bm kn}(t)
   &=
    \int_{-\infty}^t \diff t'  G^{\rm R}_{\bm kn}(t-t')
    V^{\rm cl}_{\bm kn}(t')    \, .
\end{align}
%

Here, $V^{\rm cl}_{\bm kn} (t)$
is the inverse Fourier transform of Eq.~\eqref{eq:def_of_V}, 
giving its time-domain representation
%
\begin{align}
    V^{\rm cl}_{\bm kn} (t) = 
        \sum_{i\xi} g_{i,\bm kn}^\xi \mathscr O_i^\xi (t)  \, .
\end{align}
%
One can verify by direct differentiation that $A_{\bm kn}$
satisfies the first-order differential equation
%
\begin{align}
    \imu z_{\bm kn} \dot A_{\bm kn}
        = \omega_{\bm kn} A_{\bm kn} + V_{\bm kn}^{\rm cl}(t)   \, ,
    \label{eq:A}
\end{align}
%
where $z_{\bm kn}$ and $\omega_{\bm kn}$ are the same 
parameters appearing in the phonon Green's function 
[Eq.~\eqref{eq:GR_concrete_expr}].
Physically, each auxiliary variable $A_{\bm kn}$ represents a
dynamical phonon mode that effectively mediates interactions 
between electrons.
In this way, the memory integral is replaced by auxiliary 
variables, reducing the problem to a Markov process 
without explicit dependence on the full history.

\subsection{Approximation for Noise}
\label{sec:Markov-Noise}
\subsubsection{Colored Noise}

Next, we analyze the properties of the noise appearing in 
the equations of motion.  
In general, the process is non-Markovian because the noise 
has long-time correlations
%
\begin{align}
    \la \zeta_{\bm kn}^*(t) \zeta_{\bm kn}(t') \ra
    = F_{\bm kn}(t-t')  \, ,
\label{eq:noise.zeta}
\end{align}
%
where $F_{\bm kn}(t - t')$ is the Fourier transform of 
the spectral function defined in Eq.~\eqref{eq:color_spectrum}.
These time-nonlocal correlations imply that the noise is 
\textit{colored} and retains memory of past times, which 
complicates numerical integration.

Fortunately, in certain limits, the correlation function becomes local
by considering the noise in an alternative representation.  
Without loss of generality, the same correlation function is
reproduced by introducing the following auxiliary stochastic 
differential equation
%
\begin{align}
    &\imu z_{\bm kn} \dot \zeta_{\bm kn} = 
            \omega_{\bm kn}\zeta_{\bm kn} +\Gamma_{\bm kn} (t)  \, ,
    \label{eq:zeta}
\end{align}
%
where the spectrum of the newly introduced noise term 
$\Gamma_{\bm kn}$ follows from combining 
Eqs.~\eqref{eq:def_z}, \eqref{eq:Ffunc}, and 
\eqref{eq:color_spectrum}, yielding
%
\begin{align}
    &\la \Gamma^*_{\bm kn}(\ep) \Gamma_{\bm kn}(\ep') \ra =  
        \frac{ \gm_{\bm kn} \ep}{ \omega_{\bm kn} }\coth \frac{\ep}{2T} 
        \ 
        \delta(\ep-\ep')      \, .
    \label{eq:noise_type_1}
\end{align}
%
%
%
Since Eq.~\eqref{eq:zeta} closely resembles Eq.~\eqref{eq:A}, 
$\zeta_{\bm kn}$ can be interpreted as describing phonon degrees 
of freedom associated with coupling to the bath.
Note that the prefactor in Eq.~\eqref{eq:noise_type_1} is 
directly related to the imaginary part of the 
system phonon 
self-energy from coupling to the bath, 
\mbox{$- {\rm Im\,}\Pi_{\bm kn, \bm k'n'}(\ep+\imu 0^+)$}, 
defined in Eq.~\eqref{eq:Pi_simplified}.

The epLD 
equation of motion in Eq.~\eqref{eq:eom} can then be rewritten as
two coupled linear differential equations,
%
\begin{widetext}
    \begin{subequations}
    \begin{align}
        -\imu  \sum_{q} B_{pq}(\Omega_i) \dot \Omega_{iq} 
        &= \sum_{\xi} 
        \frac{\partial \mathscr O^{\xi}_i}{\partial \Omega_{ip} }  
        \qty[ 
        E_i^\xi
        -
        \sum_{j\xi'}
            I_{ij}^{\xi\xi'} \mathscr O_j^{\xi'} 
        - \sum_{\bm kn} \qty(g_{i,\bm kn}^\xi a_{\bm kn}^* 
        + {\rm c.c.}) 
        ] \, ,
        \label{eq:eom.markov.electrons}
    \end{align}
    \begin{align}
        \imu z_{\bm kn} \dot a_{\bm kn}
        =\omega_{\bm kn} a_{\bm kn} +V_{\bm kn}^{\rm cl}(t) + \Gamma_{\bm kn}(t) \, ,
        \label{eq:eom.markov.phonons}
    \end{align}
    \label{eq:eom.markov}
    \end{subequations}
\end{widetext}
%
where we have introduced the effective complex phonon variable
%
\begin{align}
    a_{\bm kn} = A_{\bm kn} + \zeta_{\bm kn}    \, ,    
            \quad a_{\bm kn} \in \mathbb C \, ,
\end{align}
%
and the complex parameter 
$z_{\bm kn}=1 + \imu\gm_{\bm kn}/\omega_{\bm kn}$
as defined in Eq.~\eqref{eq:def_z}.  
Although compact in form, Eq.~\eqref{eq:eom.markov.phonons} reduces, in the limit 
$\gamma_{\bm kn} \ll \omega_{\bm kn}$, to the familiar equation of motion of a forced, 
damped harmonic oscillator, as shown in Appendix~\ref{sec:Mech.Eq.of.Motion.Phonons}.
This correspondence ensures that the resulting dynamics are physically reasonable and 
consistent.
For completeness, the associated Fokker–Planck equation is derived in 
Appendix~\ref{sec:Fokker-Planck},  which clarifies how the system 
approaches thermal equilibrium described by the Boltzmann factor.

\subsubsection{White Noise}
\label{sec:white.noise}

The equation of motion in Eq.~\eqref{eq:eom.markov} 
contains memory effects, 
and we therefore introduce
an additional approximation.
If we explicitly restore the Planck constant $\hbar$, 
the semiclassical limit corresponds to $\hbar \to 0$.
In this limit, it is natural to consider the
high-temperature regime~\cite{Kamenev_book}. 
More precisely, the condition $\hbar \to 0$ is equivalent to 
$\hbar \omega^* \ll k_{\rm B} T$, where $\omega^*$ 
denotes a characteristic energy scale of the system, 
and $k_{\rm B}$ is the Boltzmann constant.
In this regime, the noise correlation function 
in Eq.~\eqref{eq:noise_type_1} simplifies 
in its real-time representation to
%
\begin{align}
    \la \Gamma_{\bm kn}^*(t) \Gamma_{\bm kn}(t')\ra &= 
    \frac{2\gm_{\bm kn} T}{\omega_{\bm kn}}\delta(t-t') + O(T^{-1}) \, .
\label{eq:new_high_T_noise}
\end{align}
%
This expression is local in time and corresponds to a 
frequency-independent spectrum
(i.e., {\it white} noise), as in the standard Langevin description.
Consequently, the originally non-Markovian problem reduces to  
a Markovian one for both the retarded and Keldysh components, 
and the equation of motion no longer depends on the system's  
history.

\subsection{Summary of the Markov Construction}

In summary, the original equation of motion in Eq.~\eqref{eq:eom}, 
together with the colored noise defined in Eq.~\eqref{eq:color_spectrum}, 
describes a non-Markovian system with long-time memory effects 
that make numerical simulations demanding.
To overcome this difficulty, we introduced auxiliary variables that 
reformulate the dynamics of the electronic and phonon modes into a 
Markovian form.
Upon taking the high-temperature limit, the noise 
correlations reduce to local Gaussian noise.
In this formulation, the original time-nonlocal equation of motion
is replaced by the two coupled differential equations in 
Eq.~\eqref{eq:eom.markov}, driven by Gaussian noise 
$\Gamma_{\bm kn}(t)$ with correlations given 
in Eq.~\eqref{eq:new_high_T_noise}.
In the above Markov construction, no approximation has been 
employed apart from taking the high-temperature limit.

The real-valued dynamical variables $\Omega_{ip}(t)$, 
together with the complex variables $a_{\bm kn}(t)$ 
can then be propagated forward in time without reference 
to their past history.
The resulting equations are fully Markovian and straightforward 
to implement numerically, as we demonstrate in the next section.

\section{Numerical Benchmark}
\label{sec:Num.Benchmark}

In the previous section, we reformulated the general epLD  
equations of motion in Eq.~\eqref{eq:eom} as a Markov process 
in the semiclassical limit with high-temperature noise correlation 
(white noise).
In this section, we explicitly solve the equations of motion in 
Eq.~\eqref{eq:eom.markov} to benchmark both the formalism and its 
numerical implementation.

\subsection{Two-Orbital Spin Chain}

As a simple benchmark for our method, we consider a minimal multiorbital 
model that does not exhibit a finite-temperature phase transition, 
but instead shows a smooth crossover into the low-temperature regime. 
Specifically, we consider a two-orbital spin chain described by
the following microscopic Hamiltonian for the electronic degrees of 
freedom 
%
\begin{align}
    \hat{\mathscr H_e} &= -J'\sum_{i} \hat{\bm s}_{i1} \cdot \hat{\bm s}_{i2} 
                   - 2J \sum_{i}\sum_{\al=1}^2 \hat{\bm s}_{i\al} \cdot \hat{\bm s}_{i+1,\al} \, ,
\end{align}
%
where $\hat{\bm s}_{i\al}$ denotes a \mbox{spin--$1/2$} operator at 
orbital \mbox{$\alpha=1,2$} on site $i$.
We focus on the limit where the intra-site interaction $J'> 0$ is 
ferromagnetic and dominant compared to the inter-site 
interaction $J$, i.e., $J' \gg |J|$.
In this limit, spin-triplet states are energetically 
favored, 
making the triplet sector the relevant subspace 
for describing the system's low-energy physics. 
Projecting out the singlet states using the projector 
\mbox{$P_t = \prod_{i} (\frac 3 4 + \hat{\bm s}_{i1}\cdot \hat{\bm s}_{i2})$},
we obtain the effective low-energy Hamiltonian
%
\begin{align}
    \hat {\mathscr H}_e ' &= P_t \hat{\mathscr H}_e P_t 
    				= - J \sum_i \hat{\bm S}_i \cdot \hat{\bm S}_{i+1}   \, .
\label{eq:SU3.chain.electrons}
\end{align}
%
Here, $\hat{\bm S}_i$ denotes an effective spin--1 operator acting 
within the triplet subspace at site $i$.
It satisfies the spin-length relation  $\bm{S}_i^2 = S(S+1) = 2$, 
and spans a local Hilbert space with SU$(3)$ structure.

We consider localized phonons that couple to the 
time-reversal symmetric components of the electronic moments at 
each site---namely, the quadrupolar degrees of freedom
defined by  
%
\begin{equation}
    \begin{aligned}
        \hat{Q}_{3z^2-r^2} &= \tfrac{1}{\sqrt{3}} \left[ 3(\hat{S}_z)^2 - S(S+1) \right]  \, ,  
        \\
        \hat{Q}_{x^2-y^2} &= (\hat{S}_x)^2 - (\hat{S}_y)^2  \, , \\
        \hat{Q}_{xy} &= \hat{S}_x \hat{S}_y + \hat{S}_y \hat{S}_x   \, , \\
        \hat{Q}_{yz} &= \hat{S}_y \hat{S}_z + \hat{S}_z \hat{S}_y  \, , \\
        \hat{Q}_{zx} &= \hat{S}_z \hat{S}_x + \hat{S}_x \hat{S}_z  \, .
    \end{aligned}
    \label{eq:Quad.components}
\end{equation}
%
The part of the Hamiltonian that contains both the phonon energy and 
the spin-phonon coupling is given respectively by 
%
\begin{align}
	\hat {\mathscr H}_p &= \omega_0 \sum_{i\xi} \hat a_{i\xi}^\dg \hat a_{i\xi} \, , \\
	\hat {\mathscr H}_{ep} &=g \sum_{i\xi} (\hat a_{i\xi}
						+\hat a_{i\xi}^\dg) \hat Q_{i\xi}  \, ,
\label{eq:p.ep.Ham}
\end{align}
%
where $\hat{a}_{i\xi}$ denote optical Einstein phonon modes at site $i$, each 
with constant frequency $\omega_0$ (no dispersion), and five fluctuating 
components labeled by \mbox{$\xi = \{3z^2 - r^2, x^2 - y^2, xy, yz, zx \}$}. 
These phonon modes couple to the five linearly independent quadrupolar 
components in Eq.~\eqref{eq:Quad.components} 
of the electrons via the spin-phonon coupling strength $g$. 
For simplicity, we neglect any anisotropy, such that phonons have 
only one characteristic 
eigenfrequency and coupling constant. 
A similar $S=1$ model coupled to $yz$- and $zx$-type phonons 
is also studied in Ref.~\cite{Sutcliffe25}.

In this setup, the generalized epLD equations of motion from
Eq.~\eqref{eq:eom.markov},
applied to the effective spin--$1$ chain, take the explicit form
%
\begin{widetext}
\begin{subequations}
    \begin{align}
        \dot \Omega_{ip} &= \imu \sum_{q=1}^4 B^{-1}_{pq} (\Omega_i) 
        \qty[ 
    	    J \frac{\partial \bm {S} (\Omega_i)}{\partial \Omega_{iq} }  
    	    \cdot \sum_{j\in {\rm NN}} \bm {S} (\Omega_j) 
            - g \sum_{\xi} 
    	    \frac{\partial Q_{\xi}(\Omega_i)}{\partial \Omega_{iq} }  
     	   \qty(a_{i\xi} + a_{i\xi}^* )
           ] \, , 
    \label{eq:EoM.1D.model.electrons} 
    \end{align}
    \begin{align}   
        \imu z \dot a_{i\xi} &= \omega_0 a_{i\xi} + g Q_\xi(\Omega_i) 
	   					+ \Gamma_{i\xi}(t)  \, .
    \label{eq:EoM.1D.model.phonons} 
    \end{align}
    \label{eq:EoM.1D.model} 
\end{subequations}
\end{widetext}
%
These two coupled differential equations describe the real-time 
dynamics of the electronic and phononic degrees of freedom.
The damping of the phonons is captured by the complex parameter 
$z=1 + \imu\gm/\omega_0$ [see Eq.~\eqref{eq:def_z}].

We model the stochastic temperature fluctuations $\Gamma_{i\xi}(t)$ 
as white noise [see definition in Eq.~\eqref{eq:new_high_T_noise}]
%
\begin{align}
    \la \Gamma^*_{i\xi}(t) \Gamma_{j\xi'}(t') \ra = 
            \frac{
            2 \gm T}{\omega_0} \delta_{ij} \delta_{\xi\xi'} \delta(t-t')  \, .
\end{align}
%
While we here adopt white noise to mimic thermal fluctuations in their 
high-temperature limit, other forms of noise---such as colored noise---can 
in principle be implemented as well (see discussion 
in Sec.~\ref{sec:Markov-Noise}).

In numerical simulations, time is discretized as 
%
\begin{align}
    t_n = n \ \varDelta t   \, ,
\end{align}
%
with a typical time step of 
$\varDelta t = 0.01$ in units of $\hbar/J$.
The deterministic parts of Eq.~\eqref{eq:EoM.1D.model} 
are integrated using a fourth-order Runge–Kutta method~\cite{OrdinaryDiffEquations1}.
The stochastic contributions are treated using 
the Euler method, yielding the instantaneous white-noise term
%
\begin{align}
    \Gamma_{i\xi}(t_n)
    = \sqrt{\frac{\gm T}{\omega_0 \varDelta t}}
    \left(X_{i\xi n}+\imu Y_{i\xi n}\right)  \, ,
\end{align}
%
where $X_{i\xi n}$ and $Y_{i\xi n}$ are independent real-valued Gaussian 
random variables with mean $\mu=0$ and variance $\sigma^2=1$.

We represent the electronic coherent states $|\Omega_i\ra$ 
[see Eq.~\eqref{eq:def_of_c}] using the 
parametrization given in Eq.~\eqref{eq:c.alpha}~\cite{Iwazaki2023}.
Specifically, we formulate the problem in its SU(3) 
representation in order to capture all degrees of freedom 
of the effective spin--1 moments in the model of 
Eq.~\eqref{eq:SU3.chain.electrons}.
An explicit construction is provided in 
Appendix~\ref{sec:representation.SU3}.
As shown in Eqs.~\eqref{eq:param.x} and \eqref{eq:param.phi}, 
the parametrization of $c_{\alpha}(\Omega)$ is restricted to 
specific parameter domains which, in general, are not preserved 
when numerically solving the equations of motion.
Nevertheless, as we show in Appendix~\ref{sec:domain.simulations}, 
this does not pose any practical issues in simulations, since different 
domains are related by internal gauge transformations and do 
not distort the underlying physics.

\subsection{Results}

We now explicitly solve the equations of motion for the 
two-orbital spin--1 chain, 
given in Eq.~\eqref{eq:EoM.1D.model}, and 
verify that the system behaves correctly in certain limits.
The model involves five independent parameters
%
\begin{align}
    J,\ \omega_0,\  \gamma,\  g, \ T        \, ,
\label{eq:model.parameters}
\end{align}
%
where $J$ is the exchange interaction between neighboring spin--$1$ 
electronic moments, $\omega_0$ is the phonon eigenfrequency,
$\gamma$ is the phonon damping coefficient,
$g$ is the electron-phonon coupling strength, and 
$T$ is the temperature of the bath.
In the present case, both $\omega_0$ and $g$ are treated as 
wave vector $\bm k$-independent,
though they can, in general, carry momentum dependence 
(see Sec.~\ref{sec:relation.to.LLG}).

\subsubsection{Nonequilibrium Relaxation Process}
\label{sec:opt.process}

%
\begin{figure}[thb]
    \captionsetup[subfigure]{farskip=2pt,captionskip=1pt}
    \centering	
    \includegraphics[width=0.48\textwidth]{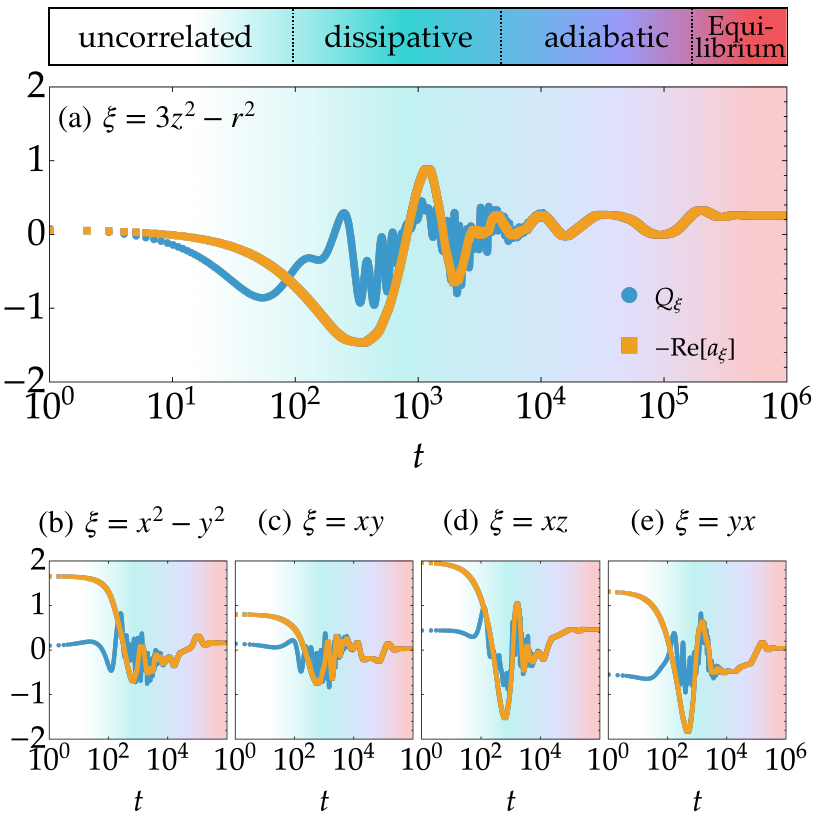}
    \caption{ 
    Numerical integration of the equations of motion in
    Eq.~\eqref{eq:EoM.1D.model} for the coupled 
    electron and phonon degrees of freedom 
    in a ferromagnetic 
    spin--$1$ chain ($J=1$) of size $N_s=100$ with periodic 
    boundary conditions at zero temperature ($T=0$).
    Shown are the expectation values of the quadrupolar moments
    $Q_{\xi}$ 
    [c.f. Eq.~\eqref{eq:Quad.components}]
    for a representative electron and its associated real-valued phonon 
    mode $-{\rm Re}\left[a_{\xi}\right]$ at site $i=0$.
    Four characteristic regimes of the dynamics can be identified:
    an uncorrelated regime ($t\approx1$), 
    a dissipative regime ($t\approx10^3$), 
    an adiabatic regime ($t\approx5\times10^4$),
    and the final equilibrium regime ($t\approx10^6$).
    Simulations are initialized from random spin and phonon configurations 
    across the chain, using parameters
    \mbox{$\omega_0 = 0.5$}, \mbox{$\gamma = 0.1$}, and 
    \mbox{$g=0.5$}. 
    The equations of motion are integrated using a discretized time 
    step $\delta t = 0.01$.
    The relaxation dynamics are further illustrated in the accompanying 
    animation~\cite{animation}, which shows a $10$-site system over the 
    first $1000$ time steps.
    }
\label{fig:benchmark.opt.quad}
\end{figure}
%

In Fig.~\ref{fig:benchmark.opt.quad} and the 
animation in~\cite{animation}, we show the real-time 
nonequilibrium relaxation 
of the electron and phonon degrees of freedom for a 
ferromagnetic spin--1 chain 
using parameters 
\mbox{$\omega_0 = 0.5$}, 
\mbox{$\gamma = 0.1$}, and 
\mbox{$g=0.5$}
at fixed bath temperature $T=0$.
Displayed are the expectation values of the 
electronic quadrupole moments $Q_{\xi}$ and corresponding 
real-components of the phonon amplitudes 
$-{\rm Re}\left[a_{\xi}\right]$ for a representative electron on 
site $i=0$.
The system is initialized in a random high-energy state 
with uncorrelated 
electronic and phononic configurations.

At early times ($t \approx 1$), the system is in an uncorrelated regime 
in which the different quadrupolar components exhibit significantly different 
behavior: the electronic moments fluctuate strongly, whereas the phonon 
amplitudes remain comparatively stable due to their lower eigenfrequency 
and finite damping.
As time progresses, the system enters a dissipative regime, where energy 
is transferred from the electronic subsystem to the phonons and subsequently 
dissipated into the thermal bath.
By $t \approx 10^4$, an adiabatic regime emerges in which the electronic 
quadrupoles become strongly correlated with the phonons and closely follow 
their motion.
Finally, around $t \approx 10^6$, the system approaches equilibrium, 
where both electronic and phononic fluctuations gradually decay as the 
system relaxes toward its ground state.

%
\begin{figure}[t]
    \captionsetup[subfigure]{farskip=2pt,captionskip=1pt}
    \centering	
    \includegraphics[width=0.48\textwidth]{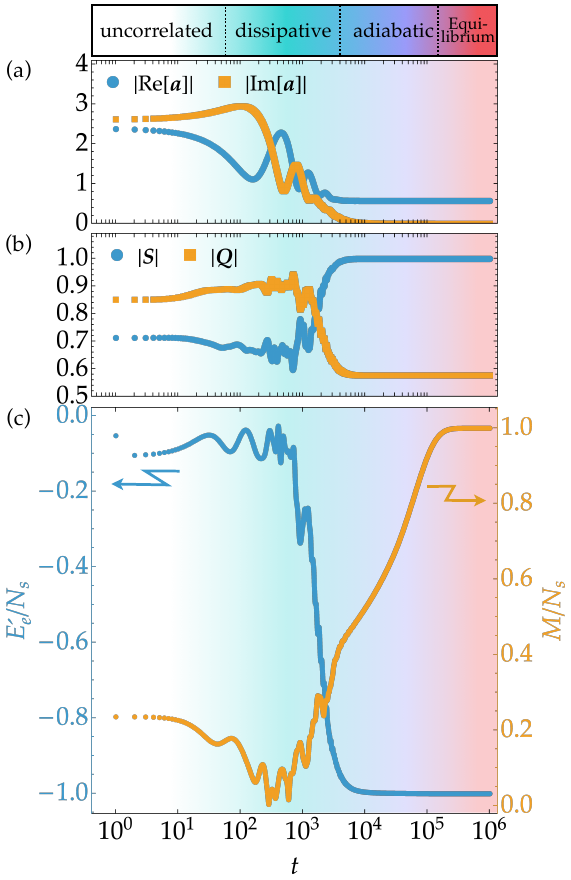}
    \caption{
    Time evolution of physical observables during numerical
    integration of the equations of motion in Eq.~\eqref{eq:EoM.1D.model}
    for the ferromagnetic ($J=1$) spin--$1$ chain, 
    using the same simulation conditions as 
    in Fig.~\ref{fig:benchmark.opt.quad}.
    Plotted are 
    (a) the norms of the phonon amplitudes, separated into 
    real $|{\rm Re}[{\bm a}]|$ [Eq.~\eqref{eq:aRe.norm}] and 
    imaginary $|{\rm Im}[{\bm a}]|$ [Eq.~\eqref{eq:aIm.norm}] components;
    (b) the norms of the electronic dipole $|{\bm S}|$ [Eq.~\eqref{eq:S.norm}]
    and quadrupole $|{\bm Q}|$ [Eq.~\eqref{eq:Q.norm}] moments;
    (c) the electronic energy per site 
    $E_e'/N_s = \langle{\hat{\mathscr H}}_e'\rangle/N_s$ 
    [Eq.~\eqref{eq:SU3.chain.electrons}] and the magnetization per site 
    $M/N_s$ 
    [Eq.~\eqref{eq:magnetization}], with all quantities plotted as functions of simulation time steps $t$.
    }
\label{fig:benchmark.opt.energy.mag}
\end{figure}
%

In Fig.~\ref{fig:benchmark.opt.energy.mag}, we show the time 
evolution of  
physical observables obtained from the same simulation shown in 
Fig.~\ref{fig:benchmark.opt.quad}.
Figure~\ref{fig:benchmark.opt.energy.mag}(a) displays the norms 
of the phonon amplitudes, separated into their real and imaginary 
components, defined as
%
\begin{align}
    \big|{\rm Re}[{\bm a}]\big| &= 
            \frac{1}{N_s} \sum_i \big|{\rm Re} [{\bm a}_i]\big|   \, ,  
    \label{eq:aRe.norm} \\
    \big|{\rm Im}[{\bm a}]\big| &= 
            \frac{1}{N_s} \sum_i \big|{\rm Im} [{\bm a}_i]\big|   \, .
    \label{eq:aIm.norm}
\end{align}
%
Figure~\ref{fig:benchmark.opt.energy.mag}(b) shows the norms of 
the electronic dipole and quadrupole moments, defined by
%
\begin{align}
    \big|{\bm S} \big| &= 
            \frac{1}{N_s} \sum_i \big|{\bm S}_i\big|   \, ,  
    \label{eq:S.norm} \\
    \big|{\bm Q} \big| &= 
            \frac{1}{N_s} \sum_i \big|{\bm Q}_i\big|   \, .
     \label{eq:Q.norm}
\end{align}
%
In addition, we plot the normalized electronic energy, 
$E_e'/N_s = \langle{\hat{\mathscr H}}_e'\rangle/N_s$
[Eq.~\eqref{eq:SU3.chain.electrons}] 
and the magnetization 
%
\begin{align}
    m_{\xi} &=  \sum_i S_{\xi}(\Omega_i)    \, , \\
    \frac{M}{N_s} &= \frac{1}{N_s} 
                           \sqrt{\sum_{\xi} m_{\xi}^2 }    \, ,    
    \label{eq:magnetization}  
\end{align}
%
with $\xi = x, y, z$. 
The explicit form of the spin-dipole expectation 
values $S_\xi(\Omega_i)$ in their SU(3) representation 
is given in Eq.~\eqref{eq:chain.dipole} of 
Appendix~\ref{sec:representation.SU3}.

The initial values of $E_e'$, $M$, and the norms of 
the electronic and phononic variables reflect the uncorrelated
high-temperature configuration from which the simulation was initialized.
During the early stage of the dynamics ($t \lesssim 10^3$), both 
the energy and magnetization exhibit strong fluctuations, 
accompanied by a noticeable drop in magnetization.
In this transient regime, the system appears to enhance the local electronic 
quadrupole moments $|Q|$, possibly as a mechanism to facilitate energy 
dissipation through more efficient coupling to phonons.
This behavior is also reflected in the increased phonon 
amplitudes during this period.

Between $t \approx 10^3$ and $10^4$, in the dissipative regime,
the energy rapidly relaxes 
and approaches the expected ground-state value 
$E_e' / N_s = -1$.
Following this relaxation, the system evolves toward a 
configuration dominated by local dipole moments with $|{\bm S}| = 1$, 
while the phonon amplitude stabilizes: the real part reaches a 
constant value and the imaginary part vanishes.
This indicates that the phonons no longer carry momentum and move 
only slowly, while being adiabatically followed by the electrons 
(see also the discussion in Fig.~\ref{fig:benchmark.opt.quad}).
Although the electronic states are fully dipolar at this point, the 
total magnetization remains below saturation and continues to 
evolve more slowly.
It reaches the fully polarized equilibrium ground-state value $M/N_s = 1$ only 
after $t \approx 10^5$ time steps, highlighting a clear separation 
of time scales between local relaxation and global magnetic ordering.
%

\subsubsection{Thermodynamics}

%
\begin{figure*}[t]
    \centering	
    \includegraphics[width=0.98\textwidth]{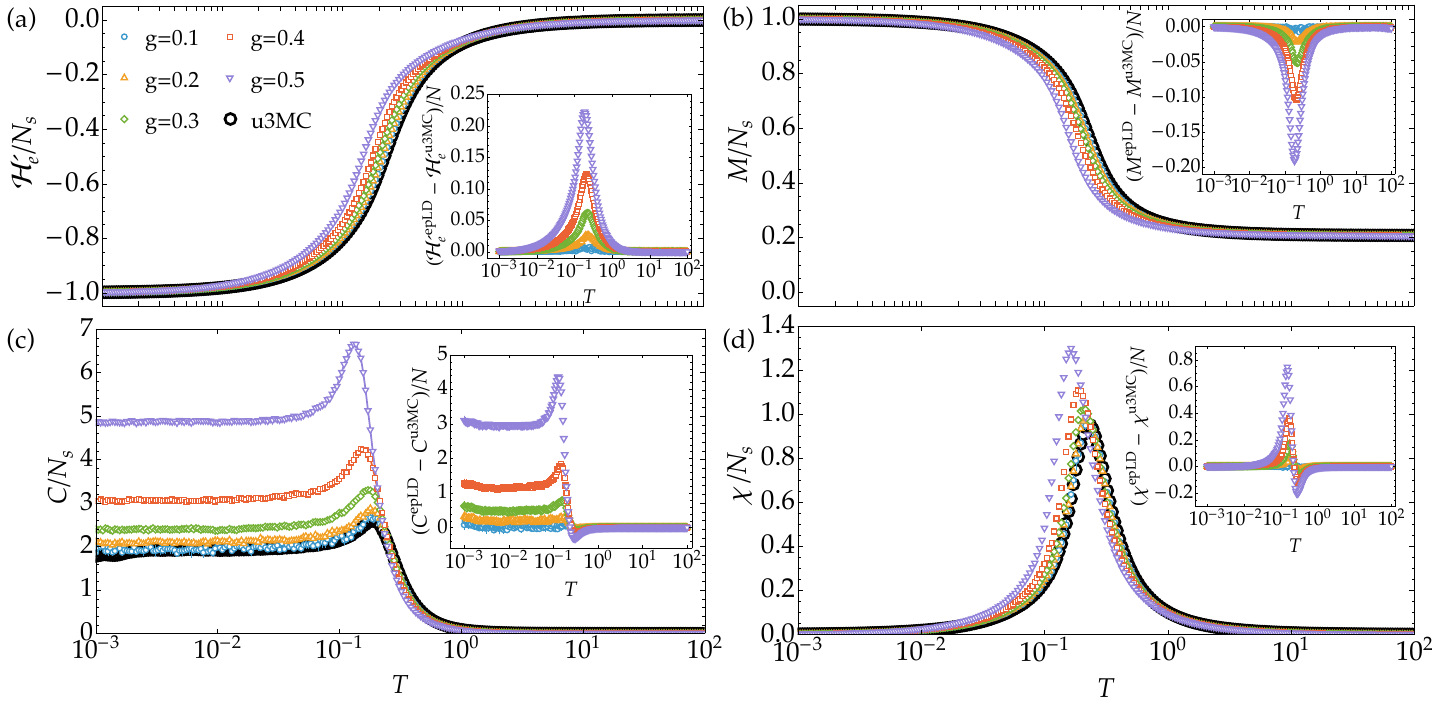} 
    \caption{
    Temperature dependence of thermodynamic observables 
    in the ferromagnetic \mbox{($J=1$)} spin--$1$
    chain with system size $N_s=10$.
    We compare results obtained using 
    SU(3) electron-phonon Langevin dynamics (epLD)  
    for various values of the electron-phonon coupling $g$, against stochastic 
    U(3) Monte Carlo (u3MC) simulations 
    (see Ref.~\cite{Remund2022} for details), which do not 
    include phonon degrees of freedom.
    Shown are the normalized 
    (a) electronic energy 
    $E_e'/N_s = \langle{\hat{\mathscr H}}_e'\rangle/N_s$ 
    [Eq.~\eqref{eq:SU3.chain.electrons}],
    (b) magnetization $M/N_s$ [Eq.~\eqref{eq:magnetization}],
    (c) specific heat $C/N_s$, [Eq.~\eqref{eq:specH}], and
    (d) magnetic susceptibility $\chi/N_s$, [Eq.~\eqref{eq:sus}].
    Insets display the differences between the epLD and u3MC results, 
    illustrating that the epLD formalism recovers the u3MC values 
    in the limit $g \to 0$.
    All epLD simulations were performed for parameters 
    $\omega_0 = 0.5$, 
    $\gamma = 0.1$, using a discretized time step 
    $\varDelta t = 0.01$.
    Simulations were initialized from random high-temperature 
    configurations, thermalized for $2\times10^6$ time steps, and 
    averaged over $10^5$ statistically independent samples. 
    Error bars were estimated from five independent simulation runs.
    }
\label{fig:benchmark.energy.temp}
\end{figure*}
%

To validate our approach at finite temperatures, we compare results 
from the 
electron-phonon-coupled Langevin dynamics (epLD) for SU(3) 
with those obtained from an independent stochastic sampling method, 
namely the U(3) Monte Carlo (u3MC) framework developed in 
Ref.~\cite{Remund2022} for spin-1 magnetic systems.
Despite their fundamentally different formulations both 
methods must reproduce the same thermodynamic behavior in 
the absence of electron-phonon coupling. 
Indeed, in the limit \mbox{$g \to 0$}, where the coupling 
vanishes, we find full agreement between the two approaches. 
This equivalence can be understood analytically from 
the corresponding Fokker–Planck equation and is discussed 
in detail in Appendix~\ref{sec:Fokker-Planck_Boltzmann}.
However, since the epLD formalism relies on phonons to mediate energy 
dissipation, it cannot equilibrate the system directly at  
$g = 0$.
We therefore 
perform simulations at several finite values of 
$g$ and extrapolate the results 
to the $g \to 0$ limit in order to 
benchmark the consistency of our method against  
the phonon-free u3MC framework.

In Fig.~\ref{fig:benchmark.energy.temp}, we present thermodynamic 
observables for the ferromagnetic spin--1 chain of size $N_s = 10$.
In one dimension, this system does not exhibit a finite-temperature 
phase transition. 
Instead, it displays a smooth crossover from a high-temperature 
paramagnetic regime to a low-temperature state with strong 
ferromagnetic correlations.
The absence of a finite-temperature phase transition makes the 
model well suited for benchmarking, 
since equilibration proceeds smoothly and is less affected by metastability or 
critical slowing down.
We present the normalized electronic energy
$E_e'/N_s = \langle{\hat{\mathscr H}}_e'\rangle/N_s$
[Eq.~\eqref{eq:SU3.chain.electrons}] 
and
magnetization $M/N_s$ [Eq.~\eqref{eq:magnetization}], as well as 
the specific heat $C/N_s$ and 
magnetic susceptibility $\chi/N_s$, defined as
%
\begin{align}
    \frac{C}{N_s} &=  \frac{1}{N_s T^2} \left[ \langle{E}_e'^{ \ 2}\rangle 
            - \langle {E}_e' \rangle^2  \right]   \, , 
    \label{eq:specH}\\
    \frac{\chi}{N_s} &=  \frac{1}{N_s T} \left[ \langle M^2\rangle 
            - \langle M \rangle^2 \right]    \, .
    \label{eq:sus}
\end{align}
%
Here, $\langle \ldots  \rangle$ denotes an average over
statistically independent samples from the time series produced 
by the numerical integration of the equations of motion.

The epLD simulations 
are initialized from a random high-temperature 
configuration and thermalized for \mbox{$2 \times 10^6$} 
time steps at the target temperature $T$.
After thermalization, thermodynamic observables are sampled every 
$300$ steps, and averages are computed over $10^5$ statistically 
independent snapshots.
To estimate error bars, we repeat each simulation five times 
with independent seeds. 
The resulting error bars are typically smaller than the symbol 
size in the plots.
For comparison, stochastic u3MC simulations are performed 
using an equivalent protocol for thermodynamic measurements: 
$10^6$ Monte Carlo steps 
of simulated annealing followed by $10^6$ steps of thermalization. 

We observe that the electronic energy landscape is significantly 
affected by the strength of the electron-phonon coupling $g$.
Stronger coupling leads to an increase in electronic energy and 
suppresses magnetization, 
particularly around the crossover region near $T \approx 0.2$, 
which separates the high-temperature paramagnetic regime 
from the low-temperature ferromagnetically correlated state.

This shift at the inflection point of both energy and 
magnetization has a pronounced effect on their associated 
response functions.
Specifically, the susceptibility peak moves to 
lower temperatures as $g$ increases, 
reflecting a renormalization of the characteristic crossover 
scale due to the additional phonon degrees of freedom.

Furthermore, the specific heat is enhanced at low temperatures 
with increasing $g$.
Within our semiclassical treatment, the active spin degrees of 
freedom are described as classical harmonic oscillators, each 
contributing $k_B / 2$ to the specific heat in the limit 
$T \to 0$~\cite{Moessner1998b, Remund2022}.
For an isolated $S=1$ system without coupling to phonons, 
there are four degrees of freedom, yielding a low-temperature 
specific heat of $C(T\to0)/N_s = 2$.
This value is reproduced by the u3MC simulations in the 
absence of phonon degrees of freedom~\cite{Remund2022}.

For finite electron-phonon coupling $g>0$, however, 
the low-temperature specific heat increases, indicating that 
the electronic subsystem effectively acquires additional 
degrees of freedom through hybridization with phonons.
Importantly, upon extrapolation to the limit $g \to 0$ we recover
%
\begin{align}
    \lim_{g\to0} C(T\to0)/N_s = 2       \, ,
\end{align}
%
in agreement with the independent u3MC results.
This completes the benchmark of thermodynamic observables for our 
epLD simulations.

\subsubsection{Dynamics}
\label{sec:dynamics.chain}

%
\begin{figure*}[thb]
	\captionsetup[subfigure]{farskip=2pt,captionskip=1pt}
	\centering	
  		\includegraphics[width=0.98\textwidth]{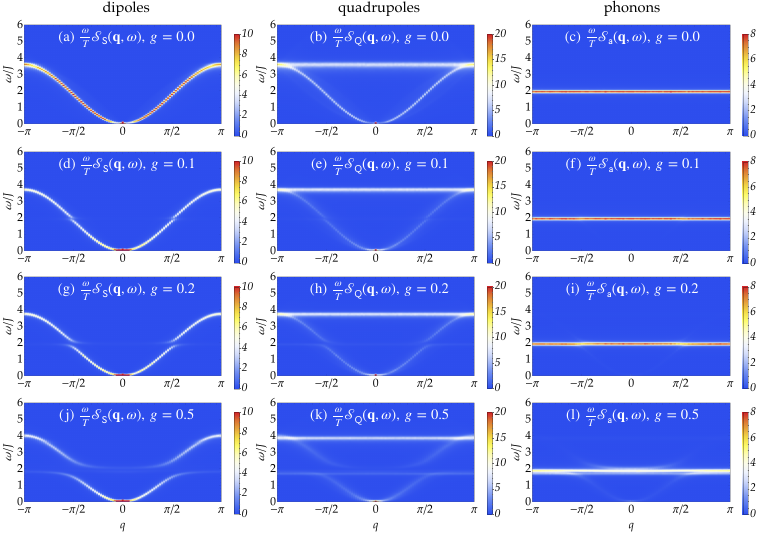} 	
    \caption{ 
    Dynamical structure factors for the ferromagnetic 
    spin--1 chain 
    ($J = 1$), obtained from numerical integration of 
    the equations of motion in 
    Eq.~\eqref{eq:EoM.1D.model},
    for parameters 
    $\gamma = 0.02$, 
    $\omega_0 = 2$, and 
    $T=0.05$.
    The panels show the dynamical structure factors 
    [Eq.~\eqref{eq:dynamical.structure.factor}]
    for spin dipoles $\tfrac{\omega}{T} \mathcal S_S(\bm q, \omega)$ (left), 
    spin quadrupoles $\tfrac{\omega}{T} \mathcal S_Q(\bm q, \omega)$ (middle), and 
    phonons $\tfrac{\omega}{T} \mathcal S_a(\bm q, \omega)$ (right), computed 
    for various electron-phonon coupling strengths $g$.
    }
    \label{fig:Sqw.FM1} 
\end{figure*}
%

%
\begin{figure*}[thb]
	\captionsetup[subfigure]{farskip=2pt,captionskip=1pt}
	\centering	
  		\includegraphics[width=0.98\textwidth]{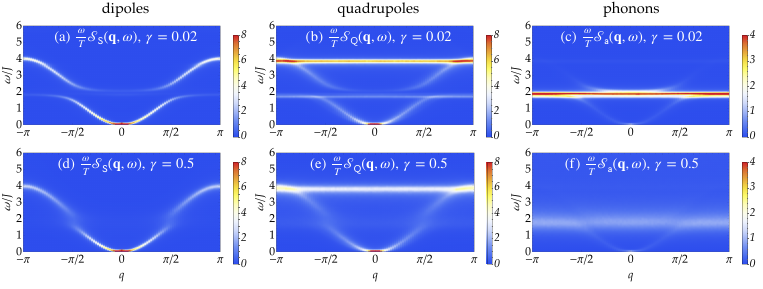} 	
    \caption{ 
    Dynamical structure factors 
    for the ferromagnetic spin--1 chain  
    ($J = 1$), corresponding to Fig.~\ref{fig:Sqw.FM1}.
    Shown are the spectra of spin dipoles 
    $\tfrac{\omega}{T} \mathcal S_S(\bm q, \omega)$ (left), 
    spin quadrupoles 
    $\tfrac{\omega}{T} \mathcal S_Q(\bm q, \omega)$ (middle), 
    and phonons 
    $\tfrac{\omega}{T} \mathcal S_a(\bm q, \omega)$ (right), computed for  
    $g = 0.5$, 
    $\omega_0 = 2$, and 
    $T=0.05$ 
    at different values of the phonon damping $\gamma$. 
    The top row reproduces the data from the bottom 
    row of Fig.~\ref{fig:Sqw.FM1} for ease of comparison.
    }
    \label{fig:Sqw.FM2} 
\end{figure*}
%

We complete our numerical benchmark by investigating the 
dynamical correlations in the ferromagnetic \mbox{spin--1}
chain.
To this end, we initialized the simulations 
from a well-optimized 
ferromagnetic ground state, obtained by solving the equations 
of motion in Eq.~\eqref{eq:EoM.1D.model} with an intermediate 
damping $\gamma = 0.1$, 
which enables the system to reach an equilibrated 
initial condition within a reasonable computational time
(see description in Fig.~\ref{fig:benchmark.opt.quad}).
Starting from this optimized state, we then solved the 
equations of motion at $T=0.05$, 
using a phonon frequency $\omega_0 = 2$
and much weaker damping $\gamma = 0.02$ 
in order to preserve the coherence of the excitations.

After sufficient equilibration, we obtain a 
time series of the electronic 
dipole ${\bm S}_i(t)$, 
quadrupole ${\bm Q}_i(t)$, 
and phonon amplitudes ${\bm a}_i(t)$,
for all sites $i$ at positions ${\bm R}_i$ on the lattice. 
These data were transformed from the real-space and 
time domain into momentum and frequency domain using a 
Fast Fourier Transform (FFT)~\cite{FFTW3}
%
\begin{equation}
    \lambda_{\xi}({\bm q}, \omega) 
		= \frac{1}{\sqrt{N_t N_s}} \sum_{i}^{N_s} \sum_{n}^{N_t}  
			e^{\imu {\bm q}\cdot{\bm R}_i} e^{\imu  \omega t_n}
            \lambda_{i \xi}(t_n)  \, ,
\end{equation}
%
where $\lambda = \{ S, Q, a \}$.
$N_s$ and $N_t$ 
denote the total number of lattice sites and the total number 
of time steps, respectively.
To minimize numerical artifacts such as the Gibbs phenomenon 
and to mimic finite frequency resolution, we convoluted the 
numerical data with a Gaussian envelope~\cite{MathematicalPhysics}.
The dynamical structure factors for all three channels were 
then obtained as
%
\begin{eqnarray}
	\mathcal{S}_{\lambda}({\bf q},\omega) = \left\langle \sum_{\xi}
		 	| \lambda_{\xi}({\bf q}, \omega) |^2 \ \right\rangle 
        \, .
\label{eq:dynamical.structure.factor}
\end{eqnarray}

In Fig.~\ref{fig:Sqw.FM1}, we show the dynamical structure 
factor for 
spin dipole,
spin quadrupole,
and phonon channels, 
after multiplying the 
prefactor $\omega/T$ to correct for classical statistics, 
following the discussions in 
Refs.~\cite{Zhang2019, Remund2022, Dahlbom2024b, Kim2025}.
In the first row of Fig.~\ref{fig:Sqw.FM1} [(a)--(c)], 
we show the case of zero electron-phonon coupling 
($g=0$).
The spin-dipole spectrum in panel (a) exhibits a clear 
quadratic dispersion near the Brillouin zone center 
$q=0$ with uniform intensity, which is consistent with the 
expected behavior of a ferromagnet.
Panel (b) shows the spin-quadrupole channel. 
In addition to a remnant of the dispersive 
spin-dipole excitations,
the spectrum features a flat band 
at \mbox{$\omega/J \approx 4$}, 
corresponding to quadrupole excitations. 
This band is dispersionless, since 
the electron interactions 
in Eq.~\eqref{eq:SU3.chain.electrons}
do not include 
effective biquadratic interactions that would otherwise 
mediate dynamics between quadrupolar modes.
The phonon spectrum in panel (c) shows a flat, 
dispersionless band at \mbox{$\omega/J = 2$}, consistent 
with the fixed phonon eigenfrequency $\omega_0$,
which has no momentum dependence.

Panels (d)–(l) in Fig.~\ref{fig:Sqw.FM1} illustrate how 
the spectra evolve as the electron-phonon coupling is 
increased to \mbox{$g = \{ 0.1, 0.2, 0.5\}$}. 
As $g$ increases, the initially flat phonon band begins 
to mix with the quadrupole spectrum, resulting in 
band repulsion in the form of an anticrossing
and the formation of hybrid modes.
This hybridization inevitably affects the spin-dipole 
sector, progressively splitting the originally 
single magnon band into two distinct branches.
For strong coupling $g/J \approx 0.5$, 
the interaction is sufficient to fully separate 
the magnon-like dispersion into two distinct  
bands. 
At the same time, the phonon band -- though originally flat by 
construction -- develops a finite momentum dependence.
This emergent phonon dispersion does not originate from any
intrinsic momentum dependence in the model parameters, 
but rather arises dynamically from strong electron-phonon 
coupling.
We note that similar hybridization effects have 
previously been reported in orbitally degenerate 
correlated electron systems using a generalized spin-wave 
approach in Ref.~\cite{Nasu13}.

As discussed above, strong electron–phonon coupling together 
with weak phonon damping allows the formation of well-defined 
coherent excitations that are hybridized between electronic 
and phononic degrees of freedom.
In Fig.~\ref{fig:Sqw.FM2}, we examine how this hybridized 
excitation spectrum evolves as the phonon damping parameter
$\gamma$ is increased.
For weak damping ($\gamma=0.02$, top row), well-defined 
hybrid modes are clearly visible, including the anticrossing 
between phonon and quadrupole excitations. 
In this regime, the electronic and phononic sectors 
remain strongly coupled and form coherent mixed excitations.
In contrast, strong damping ($\gamma = 0.5$, bottom row) 
significantly broadens the spectra and suppresses the coherent 
hybridization between electronic and phononic excitations. 
As a consequence, the anticrossing structure becomes largely 
washed out, and the magnon-like dispersion gradually 
approaches the original ferromagnetic band structure.
In this regime, strong phonon damping slows down the phonon 
dynamics and reduces their ability to coherently follow the 
electronic motion.
As a result, the synchronization between electronic and 
phononic degrees of freedom is weakened 
for larger damping $\gamma$, 
and the hybridization between the two sectors is 
suppressed.

%
\begin{figure*}[thb]
	\captionsetup[subfigure]{farskip=2pt,captionskip=1pt}
	\centering	
  		\includegraphics[width=0.98\textwidth]{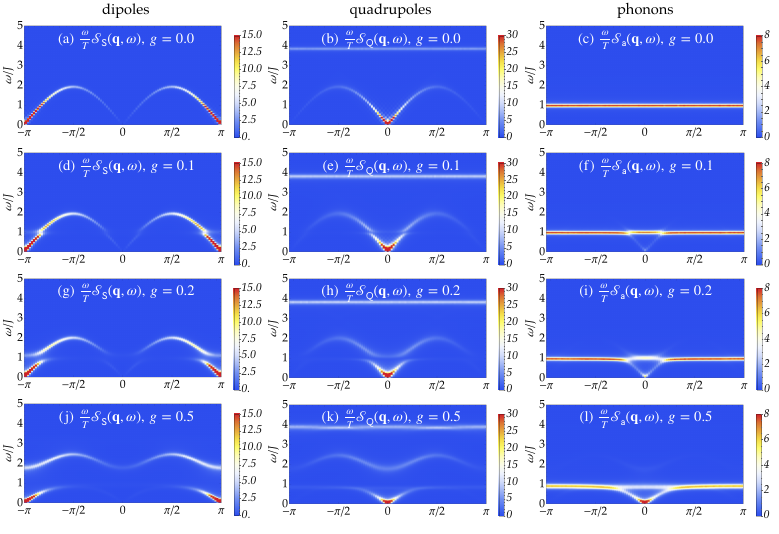} 	
    \caption{ 
    Dynamical structure factors 
    for the antiferromagnetic 
    two-orbital spin chain model ($J = -1$), 
    obtained from numerical integration of 
    the equations of motion in 
    Eq.~\eqref{eq:EoM.1D.model},
    for parameters 
    $\gamma = 0.02$, 
    $\omega_0 = 1$, and 
    $T=0.02$.
    The panels show the dynamical structure factors
    [Eq.~\eqref{eq:dynamical.structure.factor}]
    for spin dipoles $\tfrac{\omega}{T} \mathcal S_S(\bm q, \omega)$ (left), 
    spin quadrupoles $\tfrac{\omega}{T} \mathcal S_Q(\bm q, \omega)$ (middle), and 
    phonons $\tfrac{\omega}{T} \mathcal S_a(\bm q, \omega)$ (right), computed 
    for various electron-phonon coupling strengths $g$.
    }
    \label{fig:Sqw.AFM} 
\end{figure*}
%

In Fig.~\ref{fig:Sqw.AFM}, we provide 
corresponding results for the antiferromagnetic (AFM)
\mbox{spin--1} chain with $J = -1$.
Analogous to the simulations performed for the ferromagnetic 
case, we initialized the system in a well-optimized antiferromagnetic 
ground state, 
with intermediate damping $\gamma = 0.1$.
Starting from this optimized state, we then solved the equations 
of motion at $T=0.02$, using a phonon frequency 
$\omega_0 = 1$ and weak damping $\gamma = 0.02$.

Similar to the discussion of Fig.~\ref{fig:Sqw.FM1}, the first 
row of Fig.~\ref{fig:Sqw.AFM} [panels (a)--(c)] shows the case 
of zero electron-phonon coupling ($g=0$), which correctly reproduces 
the semiclassical solution of an antiferromagnet.
Panels (d)--(o) of Fig.~\ref{fig:Sqw.AFM} illustrate how the 
spectra evolve 
as the electron-phonon coupling is gradually increased to 
$g = \{ 0.1, 0.2, 0.5\}$,
illustrating the progressive hybridization between 
spin excitations and phonon modes.
For sufficiently strong coupling, the phonon modes hybridize 
strongly with the electronic degrees of freedom and acquire 
a pronounced quadrupolar character at the $\Gamma$ point ($q=0$).
As a consequence, the effective phonon modes develop a linear 
dispersion close to the $\Gamma$ point, i.e., they behave 
as emergent acoustic modes, even though the bare phonons 
in the model are dispersionless optical (Einstein) phonons.

This example clearly demonstrates how strong electron–phonon 
coupling can qualitatively modify the low-energy physics. 
Through hybridization with the electronic degrees of freedom, 
dispersionless optical phonons can evolve into effective 
dispersive modes, highlighting the substantial influence of both 
electron–phonon interactions and phonon damping on
the dynamical properties of the system.

Having benchmarked our numerical implementation and examined 
the dynamical consequences of electron–phonon coupling, we 
now turn to the connection between our approach and the widely 
used Landau–Lifshitz–Gilbert (LLG) equations, which is discussed 
in the next section.

\section{Relation to LLG equation}
\label{sec:relation.to.LLG}

%
Stochastic Landau–Lifshitz–Gilbert (LLG)
equations have been widely used to explore 
unconventional dynamical phenomena in strongly correlated 
electron systems.
It is of general interest to understand how the electron-phonon 
Langevin dynamics (epLD) for SU($N$) coherent states 
derived in this work relate to the traditional LLG equations.
In this section, we demonstrate that our epLD framework 
recovers 
the traditional LLG equations of motion in a 
specific limit when localized electrons are 
coupled to acoustic phonons.

\subsection{
Equations of Motion for Electrons Coupled to Acoustic Phonons}
\label{sec:acustic.phonons}

In the following, we focus on a form of the coupling that 
is compatible with gapless acoustic phonons at the $\Gamma$ point.
For an acoustic mode, the phonon displacement corresponds to 
a uniform translation in the limit \mbox{$\bm k \to 0$}, 
which must not couple to the electronic degrees of freedom. 
Consequently, the coupling matrix $g_{i,\bm k n}^\xi$ has 
to vanish as \mbox{$k \to 0$}.

We therefore consider a specific form of the electron-phonon system 
in which the $\bm k$-dependence of the coupling constant is assumed 
to be 
%
\begin{align}
    g_{i,\bm k n}^\xi &= 
        \frac{1}{\sqrt V} h^\xi_{n} k^{\al_n} 
        \epn^{-\imu \bm k\cdot \bm R_i}  \, ,
\label{eq:ep.coupling}
\end{align}
%
where $V$ denotes the system volume and $k = |\bm k|$.
The position of lattice site $i$ is given by $\bm R_i$,
and $h_n^\xi$ is a constant characterizing the coupling between 
the phonon mode $n$ and the electronic observable 
$\mathscr O_i^\xi$.
The corresponding acoustic phonon dispersion is
%
\begin{align}
    \omega_{\bm kn} &= c_n k     \, ,
\end{align}
%
with $c_n$ the sound velocity of mode $n$.

For simplicity, we further consider the limit 
of weak phonon damping ($\gamma_{\bm kn} \to 0$).
Performing the angular integration in $\bm k$-space and replacing the 
wave-vector sum by
\mbox{$\frac{1}{V}\sum_{\bm k} = \int \frac{\diff \bm k}{(2\pi)^3}$},
the retarded component in Eq.~\eqref{eq:retarded_expr2} becomes
%
\begin{align}
    \Sigma_{i\xi,j\xi'}^{\rm R} (t) &=
    \frac{\theta(t)}{8\pi^2 r} \sum_n h_{n}^{\xi*} h_{n}^{\xi'}
    \sum_{\sg\sg'=\pm}\sg\sg'
    \nonumber \\
    &\ \ \ 
    \times\int_0^\infty \diff k \, k^{2\al_n + 1} \, \epn^{\imu (\sg c_n k t + \sg' k r)}
    \, ,
    \label{eq:retarded_intemediate_expr}
\end{align}
%
where $r = |\bm R_i-\bm R_j|$.
The corresponding Keldysh component can be obtained from the 
fluctuation–dissipation relation in Eq.~\eqref{eq:Keldysh_def}.

Next, we specify the exponent $\al_n$. 
In Langevin dynamics, dissipative forces typically appear as 
terms proportional to the first-order time derivative of the mechanical 
variables.
In the present formalism, this corresponds to the retarded 
self-energy of the form 
$\Sigma^{\rm R}(t)\sim \delta'(t)$.
As shown below, this functional form arises for the choice
$\al_n = -1/2$.
Although this exponent does not necessarily follow from a fully 
microscopic treatment of electron–phonon coupling 
(see Appendix~\ref{sec:reaslistic_electron_phonon}), it allows us to 
recover the standard LLG equation and thus provides useful 
insight into the structure and physical interpretation of 
the generalized epLD
developed in this work

Evaluating the retarded self-energy for $\al_n = -1/2$ yields 
\begin{widetext}
%
\begin{align}
\Sigma^{\rm R}_{i\xi,j\xi'} (t-t')
= \left\{
\begin{matrix}
\displaystyle 
\sum_n \frac{h_{in}^{\xi*} h_{in}^{\xi'}}{4\pi c_n^2} \delta'(t-t')
& \ \  (i=j)
\\[5mm]
\displaystyle 
 \sum_n \frac{h_{in}^{\xi*} h_{jn}^{\xi'}}{4 \pi c_n |\bm R_i -\bm R_j|} 
\delta\qty( t-t'- \frac{|\bm R_i-\bm R_j|}{c_n} )
& \ \  (i\neq j)
\end{matrix}
\right. 
\end{align}
%
The local contribution  
($i=j$) is proportional to $\delta'(t-t')$ 
and therefore has the same structure as the Gilbert damping term.
The non-local contribution 
($i\neq j$) reflects the finite propagation 
velocity of phonons: an event at site $j$ and time $t'$ influences the 
dynamics at site $i$ after the time delay 
$|\bm R_i-\bm R_j|/c_n$.
If the phonon propagation is much faster than the localized electron 
dynamics ($c_n\to\infty$), this retardation effect can be neglected, 
and phonons effectively mediate an instantaneous interaction that 
renormalizes the interaction parameter $I_{ij}^{\xi\xi'}$.

We next consider the noise properties.
At high temperatures the Keldysh component is related to the 
retarded component through the fluctuation–dissipation relation
%
\begin{align}
    \frac{\partial}{\partial t} \Sigma_{i\xi,j\xi'}^{\rm K}(t-t')
    &= - 2\imu T \qty[ \Sigma_{i\xi,j\xi'}^{\rm R}(t-t') - \Sigma_{j\xi',i\xi}^{\rm R}(t'-t) ]  \, .
    \label{eq:relation_K_R_at_high_T}
\end{align}
%
which is derived from Eq.~\eqref{eq:Keldysh_def} at 
high temperature.

Evaluating this expression gives
%
\begin{align}
\Sigma^{\rm K}_{i\xi,j\xi'} (t-t')
= \left\{
\begin{matrix}
\displaystyle 
- \frac{\imu T}{2\pi} \sum_n \frac{h_{in}^{\xi*} h_{jn}^{\xi'}}{c_n^2} \delta(t-t')
& \ \  (i=j)
\\[5mm]
\displaystyle 
 - \frac{\imu T}{2\pi |\bm R_i - \bm R_j|}
    \sum_{n} \frac{h_{in}^{\xi*} h_{jn}^{\xi'}}{c_n}
    \theta \big( c_n |t-t'| - |\bm R_i - \bm R_j| \big)
& \ \  (i\neq j)
\end{matrix}
\right.
\end{align}
%
The local contribution ($i=j$) corresponds to white noise acting on 
each electronic site.
The nonlocal contribution represents correlated noise whose spatial 
extent is limited by the finite phonon propagation velocity 
$c_n$.

If we keep only the spatially local part as the  
dominant contribution to the electronic dynamics, the 
equation of motion reduces to
%
\begin{subequations}
\begin{align}
    -\imu  \sum_{q} B_{pq}(\Omega_i) \dot \Omega_{iq} 
    &=
    \sum_{\xi} 
    \frac{\partial \mathcal O^{\xi}_i}{\partial \Omega_{ip} }  
    \qty[ 
    E_i^\xi 
    - 
    \sum_{j\xi'}
        I_{ij}^{\xi\xi'} \mathcal O_j^{\xi'}
        - \sum_{\xi'} K_{\xi\xi'} \frac{\partial \mathcal O_i^{\xi'}}{\partial t}
        + \eta_i^\xi(t)
        ]   \, ,
    \label{eq:simplified-a}
\end{align}
\\
\begin{align}
    \big\la \, \eta_{i}^{\xi}(t) \, \eta_j^{\xi'} (t') \, \big\ra &= 2T K_{\xi\xi'} \delta_{ij}\delta(t-t') \, ,
    \label{eq:simplified-b}
\end{align}
\end{subequations}
%
\end{widetext}
where we defined 
%
$\displaystyle K_{\xi\xi'} = \sum_n 
\frac{h_{n}^{\xi*} h_{n}^{\xi'}}{4\pi c_n^2}$, 
%
and $\eta_i^\xi (t)$ is a Gaussian random  
field described by a real number.
Under these assumptions the resulting dynamics reduces to a 
Markovian Langevin equation for the electronic variables.
As shown in the next subsection, this equation can be directly 
related to the standard LLG equation.

\subsection{Comparison with LLG equation}
\label{sec:LLG.comparison}

\subsubsection{Equation of Motion for SU(2)}

For a direct comparison with the LLG equations, we now express 
the equations of motion in Eq.~\eqref{eq:simplified-a} in a 
concrete representation.
To facilitate this comparison, recall that in the LLG framework 
the classical spin moment is parameterized by two variables: the polar 
and azimuthal angles.
Accordingly, we consider the SU(2) coherent state, which is likewise 
specified by two parameters,
$\Omega = (\Omega_1,\Omega_2) = (x,\varphi)$,
following the representation introduced in 
Sec.~\ref{sec:coherent.states}, with the
explicit formulation given in Appendix~\ref{sec:coherent_state_su2}. 
In the following, we focus on the equations of motion for the 
dynamical variables at a single site and omit the site index for 
clarity.
The intersite interaction is absorbed into 
an effective local potential, i.e., 
$ E_i^\xi 
- \sum_{j\xi'} I_{ij}^{\xi\xi'} \mathcal O_j^{\xi'} 
+ \eta_i^\xi \to E_i^\xi $,
in Eq.~\eqref{eq:simplified-a}.

First, we define the damping matrix
%
\begin{align}
    \Gamma_{pq} &= \sum_{\xi,\xi'=1}^3 K_{\xi\xi'} \frac{\partial \mathcal O^\xi}{\partial \Omega_p} \frac{\partial \mathcal O^{\xi'}}{\partial \Omega_q} \, .
\end{align}
%
For simplicity we assume isotropic damping,
\mbox{$K_{\xi\xi'} = \gm \delta_{\xi\xi'}$}.
The damping matrix is then evaluated as
%
\begin{align}
\Gamma &= \gm 
\begin{pmatrix}
    4 & 0 \\
    0 & \sin^2 2x
\end{pmatrix}     \, .
\end{align}
%
We now rewrite 
the equation of motion in Eq.~\eqref{eq:simplified-a} as
%
\begin{align}
\imu \sum_q B_{pq} \dot \Omega_q = \partial_p \mathcal H + \sum_q \Gamma_{pq}\dot \Omega_q \, ,
\end{align}
%
where the effective Hamiltonian is given by
%
\begin{align}
    \mathcal H = - \sum_{\xi=1}^3 E^\xi \mathcal O^\xi  \, .
\end{align}
%

The interaction and noise contributions are not shown explicitly
but are included in the effective single-site energy $E^\xi$.
Writing the equations explicitly yields
%
\begin{align}
   \dot \varphi \sin 2x  &= 2 
    \big( E^3 \sin 2x + E^1 \cos2x\cos\varphi 
    \nonumber \\
    & \hspace{10mm} 
    + E^2\cos 2x \sin \varphi \big) - 4\gm \dot x \, ,
    \label{eq:epLD-SU2a}
    \\
    \dot x \sin 2x &= - \sin 2x
    (-E^1 \sin\varphi + E^2 \cos \varphi)
    + \gm \dot \varphi \sin^2 2x \, .
    \label{eq:epLD-SU2b}
\end{align}
%
These equations will be compared with the LLG equation in the 
next subsection.

\subsubsection{LLG Equation}

For comparison with the equations derived in the previous subsection, 
we briefly review the
LLG equation describing 
the dynamics of the magnetization $\bm M$:
%
\begin{align}
\frac{d\bm M}{dt} &= -\gm_s \bm M \times \bm H
+ \al_s \bm M \times \frac{d\bm M}{dt} \, ,
\end{align}
%
where $\gm_s$ denotes the 
gyromagnetic ratio and $\al_s$ the Gilbert 
damping coefficient.
Interactions and noise contributions may be incorporated into an 
effective magnetic field $\bm H$, in which case the equation is 
commonly referred to as the stochastic LLG equation 
\cite{Aharoni_book,Aron14,Aron16}.

The magnetization vector can be parametrized
by the polar and azimuthal angles $\theta \in [0,\pi]$ and $\phi \in [0,2\pi)$,
%
\begin{align}
    \bm M = (M_x,M_y,M_z) 
    = (\sin\theta\cos\phi, \sin\theta\sin\phi, \cos\theta)  \, ,
\end{align}
%
with the constraint $|\bm M|=1$.
Using this representation, the LLG equation can be written as
%
\begin{align}
\sin\theta \dot \phi &= \gm_s (H_z \sin \theta - H_x \cos\theta \cos \phi - H_y \cos \theta \sin\phi)
+ \al_s \dot \theta \, ,
\\
\sin \theta \dot \theta
&= \gm_s \sin \theta ( - H_x\sin \phi + H_y \cos \phi )
- \al_s \sin^2 \theta \dot \phi \, .
\end{align}
%
Comparing these equations with those obtained in the previous 
subsection [see Eqs.~\eqref{eq:epLD-SU2a} and \eqref{eq:epLD-SU2b}], 
we find a one-to-one correspondence between the 
SU(2) epLD formulation and the stochastic LLG equation:
%
\begin{align}
    \begin{matrix}
      \text{\bf SU(2) epLD}  \\[1mm]
       \pi - 2x \\
       \varphi \\
       (E^1,E^2,E^3) \\
       2\gm
    \end{matrix}    
    \ \ 
     \longleftrightarrow
     \ \ 
    \begin{matrix}
      \text{\bf Stochastic LLG }\\[1mm]
      \theta \\
       \phi \\
       \gm_s (H_x,H_y,H_z)/2 \\
       \al_s \\
    \end{matrix}    
\end{align}
%
The formal equivalence between 
SU(2) epLD and the stochastic LLG equation supports 
the validity of our approach for describing 
the dynamics of localized electronic degrees of freedom.
We emphasize, however, 
that the SU(2) formulation cannot be applied directly 
to spin systems.
Spin is a magnetic degree of freedom and 
cannot have a site-phonon coupling in Eq.~\eqref{eq:Ham_ep_coupling}.
Consequently, for applications to spin systems the present 
framework requires a larger local Hilbert space dimension ($N \ge 3$).
If one nevertheless considers the SU(2) case, a possible physical 
realization is provided by non-Kramers doublet systems 
\cite{Hart25}, where the operators $\mathcal O^{1,2,3}$ can be 
interpreted as quadrupolar and octupolar moments.

\section{Discussion}
\label{sec:Discussion}

\subsection{Realistic Coupling to Bath}
\label{sec:Discussion_reaslistic_bath}

%
It is worth commenting on the nature of thermal contact with the heat bath in realistic experimental conditions.
When maintaining a material at a constant low temperature, various experimental configurations can be employed. For instance, the sample may be immersed in liquid nitrogen or liquid/boiling helium, or it may be mounted under vacuum in thermal contact with a cooled substrate. In all such cases, thermal coupling to the heat bath is established via an interface. 
In one- or two-dimensional systems, the entire material is typically in contact with the thermal bath, so a coupling form that preserves translational symmetry, such as that in Eq.~\eqref{eq:Pi_simplified}, is likely to be justified. 
In contrast, in three-dimensional systems, heat transfer across the interface is mediated by phonons,
making the problem of phonon transmission and reflection at the boundary of fundamental importance. This interfacial thermal resistance is known as the Kapitza resistance \cite{Pollack1969,Peterson1973,Sheard1973}. While in this work we adopt a phenomenological form for it, i.e., Eqs.~\eqref{eq:bath.phonon-ham} and \eqref{eq:Ham_pb_coupling}, we anticipate that a truly first-principles theoretical description of the nonequilibrium phenomena will become possible by properly accounting for phonon transfer between the refrigerant and the sample in accordance with the specific experimental setup.
To evaluate the material’s dependency of the Kapitza resistance, the anharmonic phonon coupling would also be important especially at room temperature and above.
This realistic system-bath coupling is one of the interesting future issues to be explored.
For instance, the heating dynamics of a device or material, as recently studied in Ref.~\cite{Zanichelli26}, can be theoretically simulated within our framework for correlated electron systems.

In this work, we consider electrons directly coupled to system phonons, which subsequently 
dissipate energy from the electronic system into the thermal bath 
through their coupling to bath phonons.
If one aims to treat system phonons far from equilibrium, it is necessary to employ a 
real-time formulation after introducing the phonon self-energy due to the coupling to the 
bath in Eq.~\eqref{eq:Pi_self_energy}, instead of the electron self-energy in 
Eq.~\eqref{eq:S.diss.Matsubara} used here.
Within this framework, one can derive coupled equations of motion for both the electrons 
and the system phonons in the presence of coupling to the environment, i.e., the bath in equilibrium.
Phonon-phonon interaction would also be
important to accurately describe phonon damping.
Although we restrict the discussion in this paper to the electronic degrees of 
freedom, such a generalization is straightforward because all physical degrees 
of freedom are treated microscopically within our framework.

\subsection{Outlook for First-Principles Integration}

A promising future direction is the integration of first-principles calculations to enable quantitative analysis of real Mott insulating materials. In this study, we have constructed our theoretical framework based on a general formalism, such that the derived equations of motion can be directly applied once the parameters 
$E_i^\xi, I_{ij}^{\xi\xi'}, \omega_{\bm kn}$, and $g_{i,\bm kn}^\xi$ are specified.

Regarding the determination of electronic parameters, downfolding techniques \cite{Imada10} based on first-principles calculations allow for the construction of effective tight-binding models within the low-energy subspace near the Fermi level \cite{Marzari97,Souza01}. Within this target space, interaction parameters can be systematically estimated using the constrained random phase approximation (cRPA) \cite{Aryasetiawan04,Solovyev05}. For Mott insulators, the previous studies \cite{Iwazaki2021,Iwazaki2023} have established general procedures to derive the effective interaction $I_{ij}^{\xi\xi'}$ and onsite energy $E_i^\xi$ for localized electrons from multiorbital Hubbard models.

As for phonons, the dispersion relation $\omega_{\bm kn}$ can be obtained from first-principles phonon calculations, and the electron-phonon coupling constant $g_{i,\bm kn}^\xi$ within the relevant low-energy window can also be estimated \cite{Nomura14}.

Concerning the coupling to a thermal bath, the damping parameter $\gm_{\bm kn}$ must be provided. In situations close to thermal equilibrium, it is sufficient to assign $\gm_{\bm kn}$ phenomenologically, as the canonical distribution is correctly reproduced. For more detailed descriptions of nonequilibrium dynamics, the coupling to the thermal bath must be explicitly modeled based on the experimental setup, as discussed in Sec.~\ref{sec:Discussion_reaslistic_bath}.

In this way, a first-principles-based approach to Mott insulators can enable 
semiclassical simulations that cover not only equilibrium properties but also nonequilibrium dynamics. Validation of the method for known materials, followed by its application to unexplored compounds, may pave the way for an efficient computational exploration of functional Mott-insulator materials.

\subsection{Effect of Quantum Fluctuations}

In this work, we derived electron-phonon coupled Langevin 
dynamics (epLD) that describe the dominant semiclassical 
trajectory [Eq.~\eqref{eq:eom}] within the full quantum partition 
function defined in Eq.~\eqref{eq:Z_def}.
In this section, we discuss possible  
quantum corrections to this approximation. 
These corrections can be broadly classified into the 
following three categories:
%
\begin{enumerate}
    \item Higher-order contributions of the quantum component 
          $\Omega^{\rm qm}$ in the electronic variables of
          Eqs.~\eqref{eq:expand.cl} and \eqref{eq:expand.qm},
    \item Quantum corrections to the Berry phase term in 
          Eq.~\eqref{eq:def_Berry_curvature}, and
    \item Quantum corrections to the noise term in 
          Eq.~\eqref{eq:noise_type_1},
\end{enumerate}
%
each of which is discussed in more detail below.

1. Quantum corrections originate from the treatment of the electronic variables themselves. 
In the present formulation, the quantum component $\Omega^{\rm qm}$ introduced in Sec.~\ref{sec:real_time_keldysh} was truncated at the lowest order. 
More generally, higher-order contributions may become important, particularly in the presence of anharmonicities in the effective potential. 
In the language of conventional single-particle quantum mechanics, such corrections correspond to effects associated with anharmonic potentials and quantum tunneling~\cite{Kamenev_book}. 
As a consequence, the regime in which the quantum component becomes relevant can expand, making both the theoretical description and numerical simulations more challenging.

2. A distinct quantum correction arises from the Berry phase term. 
The parameters characterizing the coherent state of localized electrons are 
subject to the curvature of their phase space, which is associated with the Berry 
curvature defined in Eq.~\eqref{eq:def_Berry_curvature}. 
This is analogous to the case of classical spins, which are represented by 
coordinates on a unit sphere, i.e., the polar and azimuthal angles. 
Remarkably, such geometric effects,
manifest in the first term of Eq.~\eqref{eq:S0_with_real_time},
are a direct consequence of the coherent-state formulation and have no 
counterpart in the canonical phase-space structure of ordinary particle mechanics, 
where the kinetic term takes the form $p\dot{q}$ in terms of generalized coordinates and momenta.
Importantly, these correction are independent of the microscopic details of the 
system and instead reflect the choice of variables representing the physical 
degrees of freedom. 
When interpreted as the motion of a particle on a curved 
phase space, it suggests intriguing connections to geometric quantities such as 
the quantum geometric tensor~\cite{Torma23}, which has recently attracted considerable attention.

3. Last but not least, quantum corrections to the noise term can be incorporated by relaxing the high-temperature approximation used in deriving the Langevin equation as a Markovian process. In Sec.~\ref{sec:white.noise}, we assumed a high-temperature limit and obtained the white-noise correlation function in Eq.~\eqref{eq:new_high_T_noise}. 
At lower temperatures, however, quantum statistics lead to history dependent colored noise [Eq.~\eqref{eq:noise_type_1}], reflecting memory effects associated with the underlying quantum bath. 
Such corrections are well known to provide an accurate description of the dynamics of quantum particles in harmonic potentials~\cite{Schmid82,Weiss_book}.

A systematic understanding of these corrections provides a clearer decomposition 
of quantum effects in systems of localized electrons, thereby deepening our 
insight into their quantum nature.

\subsection{Simulations of Microscopic Model Hamiltonians}

The epLD formalism introduced in this work provides a 
powerful framework for studying dissipation and nonequilibrium 
dynamics in magnetic systems with arbitrary spin $S$.
Although the present formulation does not account for long-range 
quantum entanglement beyond a single site, it fully captures 
the local Hilbert space structure of electrons in terms of 
SU($N$) coherent states, correctly describing all local 
multipoles (dipole, quadrupole, etc.).
This makes our approach ideally suited for large-scale 
simulations of microscopic model Hamiltonians, allowing for 
systematic investigations of finite-size effects and direct 
comparisons with experimental results.

As a first step, it would be natural to go beyond the benchmark 
example of the spin--1 chain used in Sec.~\ref{sec:Num.Benchmark} 
and apply the formalism to higher-dimensional lattices. 
These systems can exhibit finite-temperature phase transitions, 
including continuous, first-order, or topological transitions,
which may be studied in the presence of dissipation.
A particularly promising direction involves the study of spin 
nematic phases, which host a variety of topological excitations. 
Their nonequilibrium dynamics--especially in response to external 
perturbations--are of great interest for potential applications 
in spintronic and quantum information devices.

Another intriguing avenue is the interplay between lattice dynamics 
and classical spin-liquid behavior. 
Spin liquids are known to exhibit strong dynamical signatures 
even at very low temperatures, and the effect of phonons on 
their stability remains an open question.
A notable example is the $S=1$ pyrochlore magnet 
NaCaNi$_2$F$_7$~\cite{Plumb2019, Zhang2019}, 
which has been suggested to relate to a RP$^2 \times$ U(1) 
spin liquid stabilized by large negative biquadratic 
interactions~\cite{Pohle2025}.
Such interactions imply significant coupling to lattice degrees 
of freedom~\cite{Gao2024} and call for further investigation using the present 
formalism.

Previous studies on spin liquids investigated their 
stability in the presence of bond-phonon coupling~\cite{Ferrari2021, Ferrari2024}. 
However, other regimes --- particularly those involving 
site-phonon coupling --- may instead have a stabilizing effect.
While in multiorbital systems bond phonons are 
expected to be
energetically less favorable than site phonons [see 
discussion around Eq.~\eqref{eq:bond-phonon.energy}], 
singlet formation on bonds can, 
in certain limits, be effectively described using an 
SU(4) framework~\cite{Dahlbom2024b}, 
offering a potential route to incorporating bond phonons 
into this formalism.

Because the equations of motion [Eq.~\eqref{eq:eom.markov}] 
are general, external
magnetic or electric fields can be incorporated directly, enabling 
simulations of driven systems and exotic 
topological solitons such as 
CP$^2$ skyrmions~\cite{Amari2022, Zhang2023}, 
or defect formation processes described by the 
Kibble–Zurek mechanism~\cite{Zurek2005}.
This opens up the possibility of realistically modeling pump–probe 
experiments, where energy dissipation and relaxation occur 
through microscopically grounded damping mechanisms.

\subsection{Extension to fermionic systems}

Although in this paper we have focused on localized electrons in Mott insulators, we expect the epLD framework to be applicable to itinerant correlated electron systems by employing the unconventional coherent-state representation~\cite{Yamasaki26}, in which multiorbital fermionic coherent states are constructed locally under the principle of minimizing the number of Grassmann degrees of freedom. 
The resulting basis functions contain only a single species of Grassmann variable at each site and share the same SU($N$) coherent-state structure as in the present formulation. 
This structure enables the application of a semiclassical approximation, which substantially reduces the computational cost and makes simulations of nonequilibrium systems feasible. 
Despite this advantage, an explicit epLD formulation involving fermionic degrees of freedom (i.e., Grassmann variables) remains an important challenge for future work.

\section{Summary}
\label{sec:Summary}

We have developed a microscopic framework for spin-orbital coupled
Mott insulators with electron-phonon interactions by deriving 
generalized equations of motion from a Kugel-Khomskii-type 
Hamiltonian. 
Using SU($N$) coherent states within the Keldysh path-integral 
formalism, our approach systematically incorporates both 
dissipation and stochastic fluctuations arising from coupling 
to a phonon bath. 
This leads to semiclassical equations of motion featuring 
microscopically derived damping and noise terms--providing a 
significant improvement over traditional phenomenological models.

We benchmark the formalism numerically using a two-orbital 
spin chain coupled to Einstein phonons, and demonstrate the 
emergence of hybridized electron-phonon excitations. 
In a specific limit, our equations reduce to the well-known 
Landau-Lifshitz-Gilbert (LLG) form, establishing a direct 
connection to existing spin dynamics models.

Overall, our approach enables realistic simulations of both 
equilibrium and nonequilibrium dynamics in strongly correlated 
systems, and offers a promising route to bridge first-principles 
band-structure calculations with microscopic dynamical behavior.
Beyond the present benchmark calculations, the framework can be 
extended to incorporate more realistic bath geometries, nonequilibrium 
phonon dynamics, and microscopic parameters obtained from 
first-principles methods. 
More broadly, it provides a versatile platform for investigating 
dissipative nonequilibrium phenomena in correlated systems, including 
phase transitions, topological defect dynamics, and 
frustrated magnetic systems.d

\begin{acknowledgements}

The authors are pleased to acknowledge fruitful discussions with 
Yutaka Akagi, 
Cristian Batista, 
Ryuta Iwazaki,
Nic Shannon,
and
Philipp Werner.
This work was supported by 
KAKENHI Grants 
No.~25K17335,
No.~24K00578, and No.~23K25827, 
MEXT as "Program for Promoting Researches on the Supercomputer Fugaku"
(Grant \mbox{No. JPMXP1020230411}),
and the Grant-in-Aid for Transformative Research Areas (A) 
“Correlation Design Science” (KAKENHI Grant 
No. JP25H01247, 
No. JP25H01249, 
No. JP25H01252) 
from JSPS of Japan.
Numerical calculations were carried out using HPC facilities provided by 
the Supercomputer Center of the Institute for Solid State Physics, 
the University of Tokyo.

\end{acknowledgements}

%
%
\appendix									

\section{Relation to Mechanical Equation of Motion}
\label{sec:Mech.Eq.of.Motion.Phonons}

In Sec.~\ref{sec:NonMarkov.to.Markov}, 
we derived the equations of motion 
for localized electrons coupled to phonons.
As a result, we obtained 
differential equations for electrons 
$\Omega$ [Eq.~\eqref{eq:eom.markov.electrons}] and 
phonon-like degrees of freedom $a$ 
[Eq.~\eqref{eq:eom.markov.phonons}].
In order to better understand the physical meaning 
of $a$, we compare it to the quantum-mechanical 
equation of motion for phonons in absence of
temperature fluctuations ($\Gamma$) and 
dissipation effects ($\gamma$).
Then the Heisenberg equation of motion for phonons in 
operator form gives
%
\begin{align}
    \imu \partial_t  \hat a_{\bm kn}(t) 
    &= \omega_{\bm kn} \hat a_{\bm kn}(t) 
      + \sum_{i\xi} g_{i,\bm kn}^\xi \hat{\mathscr O}_i^\xi(t)   \, ,
\label{eom.phonons.operator}    
\end{align}
%
where $\hat{\mathscr O} (t) = \epn^{\imu (\mathscr H_e + \mathscr H_p  +\mathscr H_{ep} )t } \hat{\mathscr O} \epn^{ - \imu (\mathscr H_e + \mathscr H_p  +\mathscr H_{ep} )t }$ denotes the Heisenberg picture of the operator.
This equation of motion is -- once the annihilation operator 
$\hat{a}$ is replaced by a complex number -- identical to  
Eq.~\eqref{eq:eom.markov.phonons} 
in the absence of coupling to the bath.

Furthermore, we point out that Eq.~\eqref{eq:eom.markov.phonons} 
takes the form of a forced damped harmonic oscillator.
To see this explicitly, we rewrite it in simplified form  
%
\begin{align}
    \imu z \dot a = \omega a - f        \, ,
\label{eq:eom.simple}
\end{align}
%
where we have omitted the index $\bm kn$ for clarity.
The term $f$ represents the
force from electron-phonon coupling and thermal noise. 
We now define real mechanical variables for phonons
%
\begin{align}
    a = q+\imu p \, ,
\end{align}
%
where $q$ and $p$ respectively correspond to displacement 
and momentum of a harmonic oscillator.
Substituting this into the equation of motion 
[Eq.~\eqref{eq:eom.simple}]
yields two coupled second-order differential equations
%
\begin{align}
    & \qty(1+\frac{\gm^2}{\omega^2}) \ddot q 
    = -\omega^2 q - 2\gm \dot q+\omega {\rm Re\,}f 
    +\frac{\gm}{\omega} {\rm Re\,}\dot f 
    - {\rm Im\,} \dot f   \, ,
    \label{eq:semiclassical_position}
    \\
    & \qty(1+\frac{\gm^2}{\omega^2}) \ddot p 
    = -\omega^2 p - 2\gm \dot p + \omega {\rm Im\,}f 
    + \frac{\gm}{\omega} {\rm Im\,} \dot f 
    + {\rm Re\,}\dot f   \, .
\end{align}
%
In the limit $\gm\ll \omega$, these reduce to the familiar 
equations of motion for a forced damped harmonic oscillator.
This confirms that our formalism captures the expected mechanical 
behavior of phonons coupled to both electrons and a dissipative bath.

Using a path-integral formalism for local phonons, a semiclassical equation of motion for phonons has been derived in Ref.~\cite{Picano23-1}, which is second-order in the time derivative, as in standard harmonic oscillators. In contrast, our equation of motion is obtained by reformulating the purely-electronic equation of motion and involves a first-order time derivative with complex mechanical variables ($a_{\bm kn}$).

\section{Fokker-Planck Equation}
\label{sec:Fokker-Planck}

\subsection{General Derivation}

Let us consider the probability distribution function by taking the average over the random force.
Unlike the Langevin equation, which tracks individual stochastic trajectories, the Fokker–Planck equation directly provides the time evolution of the probability density, making it well-suited for analyzing ensemble behavior and deriving steady-state distributions.
First of all, we write the equation of motion 
in a compact symbolic way:
%
\begin{equation}
	\begin{aligned}
		\dot \Omega_{ip} &= \mathcal F_{ip} (\bm \Omega, \bm q, \bm p)	\, ,  \\
  		\dot q_{\bm kn} &= \mathcal G^{(1)}_{\bm kn} (\bm \Omega, \bm q,\bm p) - \frac{1}{|z_{\bm kn}|^2} \qty( {\rm Im\,}\Gamma_{\bm kn}(t)- \frac{\gm_{\bm kn}}{\omega_{\bm kn}}{\rm Re\,}\Gamma_{\bm kn}(t))	\, ,  \\
 		\dot p_{\bm kn} &= \mathcal G^{(2)}_{\bm kn} (\bm \Omega,\bm q, \bm p) - \frac{1}{|z_{\bm kn}|^2} \qty( {\rm Re\,}\Gamma_{\bm kn}(t) + \frac{\gm_{\bm kn}}{\omega_{\bm kn}}{\rm Im\,}\Gamma_{\bm kn}(t))	\, .  \\
	\end{aligned} 
	\label{eq:eom_rewrite}
\end{equation}
%
Here we use a real-number representation 
of the phonon variables,
%
\begin{equation}
	\begin{aligned}
		a_{\bm kn} = q_{\bm kn} + \imu p_{\bm kn}	\, ,
	\end{aligned}
\end{equation}
%
and introduce the compact notation
\mbox{$\bm \Omega = \{ \Omega_{ip} \}$}, 
\mbox{$\bm q = \{ q_{\bm kn} \}$}, and 
\mbox{$\bm p = \{ p_{\bm kn} \}$}.
The functional forms 
$\mathcal F_{ip} (\bm \Omega, \bm q, \bm p)$ and 
$\mathcal G^{(1,2)}_{\bm kn} (\bm \Omega,\bm q, \bm p)$
represent the deterministic parts of the equations of motion and 
follow directly from Eq.~\eqref{eq:eom.markov}.

We define the probability density by
\begin{align}
    &P(\bm X , \bm x , \bm y,t)
    = 
    \delta \big( \bm X - {\bm \Omega}(t) \big)
    \delta\big( \bm x - \bm q(t) \big)
    \delta\big( \bm y - \bm p(t) \big) \, .
    \label{eq:gen_P_def}
\end{align}
The time-dependent components, such as $\Omega_{ip}(t)$ and $q_{\bm kn}(t)$, are determined associated with history of the noise configuration at every $t$ in the past under the specific initial condition.
Note that $\bm X,\bm x$ and $\bm y$ are just parameters and are not dependent on time.
Taking the time derivative of $P$, using 
Eq.~\eqref{eq:eom_rewrite}, and then averaging over the random force (see Appendix A5 of Ref.~\cite{Risken_book}), 
we obtain the Fokker-Planck equation for $W(t) = \la P(t) \ra $:
\begin{align}
    \frac{\partial W}{\partial t}
    &= \bigg[
    - \mathcal F_{ip} \frac{\partial}{\partial X_{ip}}
    - \mathcal G^{(1)}_{\bm kn} \frac{\partial}{\partial x_{\bm kn}} 
    - \mathcal G^{(2)}_{\bm kn} \frac{\partial}{\partial y_{\bm kn}} 
    \nonumber \\
    &\hspace{4mm}
    + \frac{ \gm_{\bm kn} T}{2\omega_{\bm kn} |z_{\bm kn}|^2}
    \qty( \frac{\partial^2}{\partial 
    x^{2}_{\bm kn}} 
    + \frac{\partial^2}{\partial y^{2}_{\bm kn}} )
    \bigg]W     \, ,
\end{align}
where the Einstein rule is employed for the summation with respect to the repeated indices ($i,p,\bm k,n$).
We omit the arguments of $\mathcal F$ and $\mathcal G^{(1),(2)}$ for brevity.

The probability distribution function for the electronic part
is obtained by taking the partial trace over phonons as
\begin{align}
    W_{\rm el}(\bm X,t)
    &= \int \diff \bm x \diff \bm y \ 
        W(\bm X , \bm x , \bm y ,t) \, ,
\end{align}
which describes the statistics of the electronic states at every time $t$.
The normalization condition $\int \diff \bm X W_{\rm el} = 1$ and the positivity $W_{\rm el}\geq 0$ is satisfied by the 
definition in Eq.~\eqref{eq:gen_P_def}.

\subsection{Explicit Example}
\label{sec:Fokker-Planck_Boltzmann}

Although the Fokker–Planck equation is generally analytically 
intractable, we present here an exactly solvable example for 
the damped harmonic oscillator.
Let us consider the $\gm\to 0$ limit of the equation of motion 
for phonons, without coupling to electrons
%
\begin{align}
    \dot q &= \omega p - \gm q - {\rm Im\,} \Gamma      \, ,
    \\
    \dot p &= -\omega q - \gm p + {\rm Re\,} \Gamma     \, ,
\end{align}
%
where we have omitted the index $\bm kn$.
Namely, the problem is reduced to the two-coupled differential 
equations with damping and noise.
We define
%
\begin{align}
    P(x,y,t) &= \delta \big(x-q(t)\big) \delta\big(y-p(t)\big)  \, ,
\end{align}
%
and obtain
%
\begin{align}
    \dot P &= - \partial_x (\omega y - \gm x - {\rm Im\,}\Gamma) P
    - \partial_y (-\omega x - \gm y + {\rm Re\,}\Gamma) P   \, .
\end{align}
%
Defining $W = \la P\ra$, we derive the following Fokker-Planck equation
%
\begin{align}
    &\dot W =
    \bigg[ - \partial_x (\omega y - \gm x)
    - \partial_y (-\omega x - \gm y )
    +  \frac{\gm T}{2\omega} (\partial_x^2+\partial_y^2) \bigg]W
    \nonumber \\
    &\hspace{-3mm}= 
    \bigg[ \omega (x\partial_y - y\partial_x) + \gm \partial_x \qty(x + \frac{T}{2\omega}\partial_x) + \gm \partial_y \qty(y + \frac{T}{2\omega}\partial_y) \bigg]W \, .
\end{align}
%
Hence, we confirm that
\begin{align}
    W_{\rm equib}(x,y) &= \exp[-\beta \omega (x^2+y^2)] \, ,
\end{align}
is the solution of the stationary state ($\dot W=0$).
This is simply the Boltzmann factor in the canonical ensemble,
as expected.

\section{Coherent-State Representation}

\subsection{Representation in SU(2)}
\label{sec:coherent_state_su2}

In Sec.~\ref{sec:LLG.comparison}, we discuss the relation 
of our epLD formalism to the well-known LLG equations.
To enable this comparison, we need to consider the SU(2) 
coherent state representation which is explicitly 
given for $c_\al (\Omega_i)$ in Eq.~\eqref{eq:c.alpha} as
%
%
\begin{align}
c_1(\Omega) &= \cos x  \, ,
\\
c_2(\Omega) &= \epn^{\imu \varphi}\sin x \, ,
\end{align}
%
where $x = \Omega_{1} \in [0,\pi/2]$ and $\varphi = \Omega_{2}\in [0,2\pi)$.
The Berry curvature matrix of Eq.~\eqref{eq:def_Berry_curvature}
is given by
%
\begin{align}
\imu B(\Omega) &= 
\sin 2x
\begin{pmatrix}
    0 & -1 \\
    1 & 0
\end{pmatrix} \, .
\end{align}
%
The pseudospin variables, defined in Eq.~\eqref{eq:O.coheren.operator},
are then given by
%
\begin{align}
    S^x = \mathcal O^{\xi=1} &= \sum_{\al\al'} c_\al^*\sigma^1_{\al\al'} c_{\al'} 
        = \sin 2x \cos \varphi      \, ,
    \\
    S^y = \mathcal O^{\xi=2} &= \sum_{\al\al'} c_\al^*\sigma^2_{\al\al'} c_{\al'} 
        = \sin 2x \sin \varphi   \, ,
    \\
    S^z = \mathcal O^{\xi=3} &= \sum_{\al\al'} c_\al^*\sigma^3_{\al\al'} c_{\al'} 
        = \cos 2x    \, ,
\end{align}
%
where ${\sigma}^{1,2,3} \equiv {\sigma}^{x,y,z}$ 
denote the standard Pauli matrices.
In this parametrization, $2x$ and $\varphi$ respectively correspond 
to the polar and azimuthal angles of a classical spin vector 
on the Bloch sphere.

\subsection{Representation in SU(3)}
\label{sec:representation.SU3}

For the numerical benchmark in Sec.~\ref{sec:Num.Benchmark}, 
we choose to simulate a two-orbital spin chain that, in its 
low-energy sector, can be represented by localized spin--1 
moments.
The local Hilbert space of the spin coherent state is 
therefore described by SU(3). 
Following the construction in Eq.~\eqref{eq:c.alpha}, 
the coefficients take the explicit form
%
\begin{equation}
    \begin{aligned}
	c_1(\Omega) &= \cos{x_1}		\, , \\
	c_2(\Omega)	&= e^{i\varphi_1} \cos{x_2} \sin{x_1}	\, ,	\\
	c_3(\Omega)	&= e^{i\varphi_2} \sin{x_1} \sin{x_2}	\,  .	
    \end{aligned}
\end{equation}
%
The corresponding Berry curvature matrix $B(\Omega)$, as defined in 
Eq.~\eqref{eq:def_Berry_curvature}, is given by
\begin{widetext}
\begin{align}
	\imu B(\Omega) =  \begin{pmatrix}
		0	&	0	& 	-\sin{2 x_1}\cos^2{ x_2}  	&	-\sin{2 x_1}\sin^2{x_2} 	\\
		0	&	0	& 	\sin^2{ x_1} \sin{2 x_2} 	&	-\sin^2{x_1} \sin{2x_2}	\\
		\sin{2x_1}\cos^2{x_2}  	&	-\sin^2{x_1} \sin{2 x_2}	&	0	&	0	\\
		\sin{2x_1}\sin^2{x_2}  	&	\sin^2{x_1} \sin{2 x_2}	& 	0	&	0	\\
	\end{pmatrix} \, .
\end{align}
%
By using the Gell-Mann matrices as generators of SU(3) in 
Eq.~\eqref{eq:O.coheren.operator}, we obtain the pseudospin 
variables for electronic spin-dipole moments 
%
\begin{equation}
    \begin{aligned}
	S_x 	&=  \sqrt{2} \cos{x_2} \sin{x_1} \left[ \cos{x_1} \cos{\varphi_1} +  \sin{x_1} \sin{x_2} \cos{(\varphi_1 - \varphi_2)} \right] \, , \\
	S_y 	&= \sqrt{2} \cos{x_2} \sin{x_1} \left[\cos{x_1} \sin{\varphi_1} - \sin{x_1} \sin{x_2}  \sin{(\varphi_1 - \varphi_2)}  \right] \, , \\
	S_z  &=  \cos^2{x_1} -\sin^2{x_1} \sin^2{x_2}  \, , 
    \end{aligned}
\label{eq:chain.dipole}
\end{equation}
%
and the corresponding spin-quadrupole moments  
%
\begin{equation}
    \begin{aligned}
	Q_{3z^2-r^2}  &=  \tfrac{1}{\sqrt{3}} \left[ \cos^2{x_1} - \tfrac{1}{2}\left( 1 + 3\cos{2x_2} \right) \sin^2{x_1}  \right]   \, ,  \\
	Q_{x^2-y^2}   &=  \sin{2x_1} \sin{x_2} \cos{\varphi_2}  \, , \\
	Q_{xy}  &=  \sin{2x_1} \sin{x_2} \sin{\varphi_2}   \, , \\
	Q_{xz}  &=  \sqrt{2} \cos{x_2} \sin{x_1} \left[ \cos{x_1} \cos{\varphi_1} - \cos{(\varphi_1-\varphi_2)} \sin{x_1} \sin{x_2}  \right]   \, , \\
	Q_{yz}  &=  \sqrt{2} \cos{x_2} \sin{x_1} \left[ \cos{x_1} \sin{\varphi_1} + \sin{(\varphi_1-\varphi_2)} \sin{x_1} \sin{x_2}  \right]   \, ,
    \end{aligned}
\end{equation}
%
\end{widetext}
providing all essential ingredients to evaluate the equations of motion in 
Eq.~\eqref{eq:EoM.1D.model}
of Sec.~\ref{sec:Num.Benchmark}.
In this parametrization, purely quadrupolar states with vanishing 
dipole moment are realized for 
$x_1 = \pi/4$ and $x_2 = \pi/2$, 
where the quadrupole moments lie within the $xy$ plane 
and its in-plane orientation is controlled by $2\varphi_2$.

\subsection{Domain of Variables in Coherent State}
\label{sec:domain.simulations}

We comment 
on the domains in the coherent-state variables.
The coherent state is characterized by the coefficients $c_\al(\Omega)$, 
which are expressed in terms of 
$\cos x_\al$, $\sin x_\al$, and $\epn^{\imu \varphi_{\al}}$, 
in the domains 
%
\begin{align}
        x_\al \in [0,\pi/2] \, ,  \quad  \ \varphi_\al \in [0,2\pi)  \, ,
\end{align}
%
[see Eqs.~\eqref{eq:c.alpha}--\eqref{eq:param.phi}]. 
If we change the variable as 
\mbox{$x_\al' = \pi/2 - x_\al$}, the sine and cosine 
functions are interchanged
%
\begin{align}
    \begin{matrix}
    (\sin x_\al,\cos x_\al)  &=&   (\cos x'_\al , \sin x'_\al)
    \end{matrix}    \, ,
\end{align}
%
with $x_\al' \in [0,\pi/2]$.
The representations in terms of $x_\al$ and $x_\al'$
are defined in the same domain and are equivalent.

Next, consider the following `gauge' transformations 
of the variable $x_\al$
%
\begin{align}
    x_\al &\longrightarrow x_\al^{\rm (i)} = x_\al + \pi /2 \, ,
    \\
    x_\al &\longrightarrow x_\al^{\rm (ii)} = x_\al + \pi \, ,
    \\
    x_\al &\longrightarrow x_\al^{\rm (iii)} = x_\al + 3\pi /2 \, .
\end{align}
%
Under these shifts, the sine and cosine transform as follows
%
\begin{align}
    \begin{matrix}
        \text{(i)}\ \  & (\sin x_\al, \cos x_\al)  &\to& (- \sin x_\al',   \cos x_\al') \, ,
        \\
        \text{(ii)}\ \  & (\sin x_\al, \cos x_\al)  &\to& (- \sin x_\al,   -\cos x_\al) \, ,
        \\
        \text{(iii)}\ \  & (\sin x_\al, \cos x_\al)  &\to& (\sin x_\al',  - \cos x_\al') \, ,
    \end{matrix}
\end{align}
%
resulting in sign changes of the sine and/or 
cosine functions. 
However, these extra minus signs can be absorbed by shifting 
the phase variable: $\varphi_\al' = \varphi_\al +\pi$,
where, due to the periodicity of $\epn^{\imu\varphi_\al}$, $\varphi_\al'$
can still be defined in the interval $[0,2\pi)$.
Therefore, the coherent states defined for $x_\al \in [0,\pi/2]$
are equivalent to those defined in the intervals:
%
\begin{align}
    {\rm (i)} \ \ [\pi/2,\pi], \quad 
    {\rm (ii)} \ \ [\pi,3\pi/2], \quad 
    {\rm (iii)} \ \ [3\pi/2,2\pi] \, .
\end{align}
%
Any of these representations can be used in practical calculations, 
reflecting the fact that the parametrization of coherent states is 
not unique.

\section{Realistic Electron-Phonon Coupling in Solids}
\label{sec:reaslistic_electron_phonon}

\subsection{Coupling to Acoustic Phonons}

Let us consider the electron-phonon coupling in solids.
We begin with the derivation of the electron-phonon coupling 
based on the Coulomb potential.
The electron-phonon interaction is written as
%
\begin{align}
    H &= \int \diff \bm r \, \rho(\bm r) \sum_i V(\bm r - \bm R_i)
    \, ,
    \\
    \rho(\bm r) &= \sum_\sg \psi^\dg_\sg (\bm r) \psi_\sg (\bm r)
    \, ,
\end{align}
%
where $V$ denotes 
a pseudo-potential generated by 
ions located at $\bm R_i$, including 
screening effects.
Here $\psi_\sg(\bm r)$ is the annihilation operator for an
electron at position $\bm r$ with 
spin $\sg$.

To account for lattice 
vibrations, we write the ionic position as a 
deviation from its equilibrium value, 
\mbox{$\bm R_i = \bm R_i^0 + \delta \bm R_i$}.
Expanding to linear order in the displacement yields the 
electron–phonon coupling
%
\begin{align}
    H_{ep} &= - \int \diff \bm r \rho(\bm r ) 
    \sum_i \frac{\partial }{\partial \bm r} 
    V(\bm r - \bm R_i^0) \cdot \delta \bm R_i
    \, .
\end{align}
%

The displacement operator can be 
expanded in Fourier modes as 
%
\begin{align}
    \delta \bm R_i &= \frac{1}{\sqrt {N}} \sum_{\bm kn} 
    \epn^{\imu \bm k\cdot\bm R_i^0}
    \sqrt{\frac{1}{2M \omega_{\bm kn}}}
    \ 
     \bm \epsilon_{\bm kn}  \phi_{\bm kn}  \, ,
\end{align}
%
with \mbox{$\phi_{\bm kn}=a_{\bm kn} + a^\dg_{-\bm kn}$}.
Here $M$ is the ion mass, $N$ is the number of ions, and
\mbox{$\bm \epsilon_{\bm kn}$} is the polarization vector of phonon 
mode $\bm kn$, satisfying 
\mbox{$\bm \epsilon_{\bm kn}^* = \bm \epsilon_{-\bm kn}$} 
\cite{Landau-Lifshitz-Statistical-Mechanics}.

For low-energy lattice dynamics it is convenient to employ a continuum approximation.
Introducing the unit-cell volume $v_{\rm cell}$, the Hamiltonian can be written as
%
\begin{align}
    H_{ep} &\simeq \int \diff \bm r \rho(\bm r) \int \frac{\diff \bm R}{v_{\rm cell}} \frac{\partial}{\partial\bm R} V(\bm r-\bm R)
    \nonumber \\
    &\ \ \ \cdot
    \frac{1}{\sqrt{N}} \sum_{\bm kn} \epn^{\imu \bm k\cdot \bm R} \sqrt{\frac{1}{2M\omega_{\bm kn}}} \bm \epsilon_{\bm kn}\phi_{\bm kn}
    \nonumber \\
    &= -  \frac{1}{v_{\rm cell}\sqrt{N}} \sum_{\bm kn} \imu \bm k \cdot \bm \epsilon_{\bm kn} \rho(-\bm k) V(\bm k)
    \sqrt{\frac{1}{2M\omega_{\bm kn}}} \phi_{\bm kn}
    \, .
\end{align}
%
The screened potential is expected to decay over a finite length scale, 
typically of the order of the inverse Fermi wavevector in an electron 
gas [$V(\bm k) \sim 1/(k^2+k_c^2)$].
We assume that a similar finite screening length exists in Mott 
insulators, implying that $V(\bm 0)$ remains finite.

Furthermore, we assume that the dominant contribution to the 
electron–phonon coupling arises from low-energy phonon modes.
Taking the long-wavelength limit \mbox{$\bm k\to \bm 0$} and 
focusing on the gapless longitudinal acoustic mode 
(\mbox{$\bm \epsilon_{\bm kn} = i \bm k / |\bm k|$}), we obtain
%
\begin{align}
    H_{ep} &\simeq \frac{1}{v_{\rm cell}\sqrt{N}} 
    V(\bm 0)
    \sum_{\bm kn} 
    \rho(\bm k) 
    |\bm k|  
    \sqrt{\frac{1}{2Mc_n |\bm k|}} \phi_{\bm kn}
    \, ,
\end{align}
%
where $c_n$ is the velocity of acoustic phonons.

From this expression we obtain the leading contribution 
to the coupling constant defined in Eq.~\eqref{eq:Ham_ep_coupling},
%
\begin{align}
    g_{i\bm k} \sim \sqrt{k} \   \epn^{\imu \bm k\cdot \bm R_i^0 }
    \, ,
    \label{eq:coupling_const_acoustic}
\end{align}
%
where we used the expansion
\mbox{$\rho(\bm k) \sim \sum_{i} n_i \epn^{-\imu \bm k\cdot \bm R_i^0 }$}
with the local electron density operator $n_i$ at site $i$.

This result is physically reasonable because the coupling vanishes in the 
limit $\bm k\to0$.
In this limit the lattice displacement corresponds to a uniform translation 
of the crystal, which does not generate any local deformation~\cite{Khan1984}.
For spin-orbital Mott insulators, the charge density $n_i$ is replaced by 
local electric multipole operators associated with the orbital degrees of freedom.

\subsection{Damping and Noise Spectrum}

The coupling to acoustic phonons derived in the previous 
subsection suggests the exponent $\alpha_n = 1/2$.
Evaluating the integral in Eq.~\eqref{eq:retarded_intemediate_expr} 
then yields the retarded component
%
\begin{widetext}
\begin{align}
\Sigma^{\rm R}_{i\xi,j\xi'} (t-t')
= \left\{
\begin{matrix}
\displaystyle 
- \sum_n \frac{h_{n}^{\xi*} h_{n}^{\xi'}}{4\pi c_n^4} \delta'''(t-t')
& \ \  (i=j)
\\[5mm]
\displaystyle 
- \sum_n \frac{h_{n}^{\xi*} h_{n}^{\xi'}}{4 \pi c_n^3 |\bm R_i -\bm R_j|} 
\delta''\qty( t-t'- \frac{|\bm R_i-\bm R_j|}{c_n} )
& \ \  (i\neq j)
\end{matrix}
\right.
\end{align}
\end{widetext}
%
Substituting the local contribution 
to Eq.~\eqref{eq:eom}, the damping term involves the third time 
derivative of the mechanical variables.
This implies a form of damping distinct from the conventional Gilbert damping.

Correspondingly, the Keldysh component
also requires a non-standard form, 
as it is related 
to the retarded function through 
Eq.~\eqref{eq:relation_K_R_at_high_T}.
From the local contribution we find  
that the noise spectrum is proportional to $\sim \omega^2$ which corresponds to 
colored noise.
Despite this frequency dependence, the dynamics can still be treated as 
a Markovian process when focusing on the local ($i=j$) contributions.

Finally, we emphasize that this non-Gilbert-type damping arises naturally 
for Mott insulators.
It is therefore not directly applicable to 
metallic magnets, where 
the conventional LLG equation is commonly  
used to describe  spin dynamics.

\subsection{Comment on Optical Phonons}

We briefly comment on the coupling to optical phonons. 
In Eq.~\eqref{eq:retarded_intemediate_expr}, the phonon dispersion 
\mbox{$\omega_{\bm kn} = c_n k$} appearing in the exponent 
for acoustic phonons is replaced by \mbox{$\omega_{\bm kn} = \omega_n={\rm const.}$}
for optical phonons, which is wavevector-independent 
in the long-wavelength limit.
As a consequence,  
\mbox{$\Sigma^{\rm R}(t)
\sim \epn^{\imu \omega_n t}$}, indicating 
that the dynamics 
inevitably dependent on the history of the system, i.e., it becomes 
non-Markovian. 
In such situations, numerical simulation based on a Markovian 
description can still be carried out efficiently by explicitly 
including semiclassical phonon degrees of freedom, as discussed in Sec.~\ref{sec:NonMarkov.to.Markov}.

We also note that the electron–phonon coupling 
is enhanced in the long-wavelength limit ($q\to 0$)~\cite{Verdi15,Sohier16}.
Therefore, the coupling to optical phonons is expected 
to play an important role in solid-state materials, including Mott insulators.

%
%
\bibliography{Bibliography}

\end{document}